\documentclass[usenatbib,usegraphicx,useAMS]{mn2e}
\pdfoutput=1
\usepackage{graphicx}
\usepackage{subfig}
\usepackage{verbatim}
\usepackage[usenames,dvipsnames]{color}
\usepackage{booktabs}
\usepackage{dsfont}
\usepackage{listings} 
\usepackage{amsmath}
\usepackage{soul}
\usepackage{amssymb}
\usepackage{pifont}
\usepackage{url}
\usepackage{array}
\usepackage{textcomp}

\allowdisplaybreaks[1]

\newcommand{\App}[1]{Appendix~\ref{#1}}

\newcommand{\eq}[1]{(\ref{#1})}
\newcommand{\Eq}[1]{Eq.~\ref{#1}}
\newcommand{\Eeq}[1]{Eq.~(\ref{#1})}

\newcommand{\Fig}[1]{Fig.~\ref{#1}}
\newcommand{\Sec}[1]{\S\ref{#1}}
\newcommand{\Tab}[1]{Table~\ref{#1}}

\def\ramses    {{\sc ramses}}
\def\ramsesrt  {{\sc ramses-rt}}
\def\athena    {{\sc Athena}}

\def\hei    {{\rm{He\textsc{i}}}}
\def\heii   {\rm{He\textsc{ii}}}

\def\hi    {{\rm{H\textsc{i}}}}
\def\hii   {\rm{H\textsc{ii}}}
\def\hisub {\rm{H \scriptscriptstyle I}}

\def\nh {n_{\rm{H}}}
\def\nho {n_{\rm{H,0}}}

\def\xhi {x_{\rm{H \scriptscriptstyle I}}}

\def\acc {{\rm cm} \, {\rm s}^{-2}}                         
\def\cci {{\rm{cm}}^{-3}}                                   
\def\ccg {{\rm{cm}}^{2} \, \rm{g}^{-1}}                     
\def\ccs {\rm{cm}^{3} \, \rm{s}^{-1}}                       
\def\cmmone {\rm{cm}^{-1}}                                  
\def\cs {\rm{cm}^{2}}                                       
\def\cmsqg {\rm{cm}^{2}/\rm{g}}                             
\def\emrate {\rm{photons} \; \rm{s}^{-1}}                   
\def\ergs {\rm{erg} \, \rm{s}^{-1}}                         
\def\edens {\rm{erg} \, \rm{cm}^{-3}}                       
\def\eflux {\rm{erg} \, {\rm{cm}}^{-2} \, \rm{s}^{-1}}     
\def\gcc {\rm{g} \ \rm{cm}^{-3}}
\def\gc {\rm{g} \ \rm{cm}^{-2}}
\def\hz {{\rm{Hz}}}                                         
\def\kms {\rm{km} \, \rm{s}^{-1}}       
\def\ar {{a}}       
\def\bF {{\bf{F}}}          
\def\csn {\sigma}           
\def\csH {\sigma_{\hisub}}           
\def\cred {\tilde c}      
\def\dt {\Delta t}
\def\dtrhd {\Delta t_{\rm RHD}}
\def\dx {\Delta x}
\def\ephot {{\epsilon_{\gamma}}}  

\def\Etrap {E_{\rm{t}}}
\def\fc   {{f_c}}             
\def\fEdd {f_{\rm E,V}}

\def\fyrad {f_{y,\rm{rad}}}

\def\Jacobian {{\mathcal{J}}}  
\def\kb {k_{\rm B}}
\def\kE {\kappa_{\rm E}} 
\def\kF {\kappa_{\rm F}} 
\def\kP {\kappa_{\rm P}} 
\def\kR {\kappa_{\rm R}} 
\def\ki {\kappa_i}
\def\Lbox {L_{\rm box}}
\def\lsun {L_{\rm{\odot}}}
\def\lR {\lambda_{\rm R}}
\def\momRadRate {\dot{\bf{p}}_{\gamma}}
\def\mp {m_{\rm p}}
\def\Msun {M_{\odot}}
\def\nin {n_{\rm{H,ion}}}        

\def\Prad {P_{\rm rad}}       
\def\recB {\alpha_{\rm{B}}}   

\def\rrpr {r_{\gamma}}        
\def\rs   {r_{\rm{S}}}        
\def\rS {r_{\rm{S}}}        
\def\rT   {r_{T}}             
\def\state {{\mathcal{U}}}  
\def\stromgren {Str\"omgren}
\def\tauc {\tau_{\rm c}}
\def\tauF {\tau_{\rm F}}
\def\tauV {\tau_{\rm V}}
\def\taub {\tau_{\rm box}}
\def\Tion {T_{\rm{ion}}}   
\def\Trad {T_{\rm{r}}}     
\def\trec {t_{\rm{rec}}}
\def\trun {t_{\rm{f}}}
\def\Tgas {T_{\rm{}}}      

\def\SD {D14}  
\def\KT {KT13}  
\def\RRT {R13}  

\mathchardef\mhyphen="2D

\long\def\symbolfootnote[#1]#2{\begingroup%
\def\thefootnote{\fnsymbol{footnote}}\footnote[#1]{#2}\endgroup}

\begin{document}

\title[Radiation pressure and photon diffusion]{A scheme for radiation
  pressure and photon diffusion with the M1 closure in RAMSES-RT}
\author[Rosdahl \& Teyssier] {J.~Rosdahl$^{1}$\thanks{E-mail:
    joki@strw.leidenuniv.nl} and
  R.~Teyssier$^{2}$ \\
  $^1$Leiden Observatory, Leiden University, P.O. Box 9513, 2300 RA,
  Leiden, The Netherlands \\
  $^2$Institute for Computational Science, University of Z\"urich, Winterthurerstrasse 190, CH-8057 Z\"urich, Switzerland }

\maketitle
\begin{abstract}
  We describe and test an updated version of radiation-hydrodynamics
  (RHD) in the \ramses{} code, that includes three new features: i)
  radiation pressure on gas, ii) accurate treatment of radiation
  diffusion in an unresolved optically thick medium, and iii)
  relativistic corrections that account for Doppler effects and work
  done by the radiation to first order in $v/c$. We validate the
  implementation in a series of tests, which include a morphological
  assessment of the M1 closure for the Eddington tensor in an
  astronomically relevant setting, dust absorption in a optically
  semi-thick medium, direct pressure on gas from ionising radiation,
  convergence of our radiation diffusion scheme towards resolved
  optical depths, correct diffusion of a radiation flash and a
  constant luminosity radiation, and finally, an experiment from Davis
  et al. of the competition between gravity and radiation pressure in
  a dusty atmosphere, and the formation of radiative Rayleigh-Taylor
  instabilities.  With the new features, \ramsesrt{} can be used for
  state-of-the-art simulations of radiation feedback from first
  principles, on galactic and cosmological scales, including not only
  direct radiation pressure from ionising photons, but also indirect
  pressure via dust from multi-scattered IR photons reprocessed from
  higher-energy radiation, both in the optically thin and thick
  limits.
\end{abstract}

\section{Introduction} \label{Intro.sec}

Recent years have seen great advances in the theory of galaxy
evolution, in part thanks to the insight gained from hydrodynamical
simulations. Among the clearest messages to come out of the
simulations is the necessity for feedback to regulate galaxy
evolution.  Without it, the galaxies are too massive and compact
compared to observations
\citep[e.g.][]{Suginohara:1998ig,Balogh:2001ed}. While the inclusion
of feedback from supernovae (SN) and active galactic nuclei (AGN) has
helped to relieve this so-called overcooling problem, over-compact
galaxies remain an issue in cosmological simulations \citep[][though
see \citealt{Schaye:2015gk}]{Scannapieco:2012iy}. This can partly be
traced directly to numerical overcooling, due to the lack of
resolution and/or the details of the hydrodynamical solver
\citep[e.g.][]{Creasey:2011ef, DallaVecchia:2012br, Keller:2014hk}.

Part of the problem may also be the lack of alternative feedback
mechanisms in simulations, such as cosmic rays
\citep[e.g.][]{Pfrommer:2007jj,Booth:2013fe, Hanasz:2013jb,
  Salem:2014de}, or radiation \citep[e.g.][]{1995ApJ...442..296G,
  Murray:2005jt, Krumholz:2009iu}.

Radiation feedback in particular has been employed in a number of
recent simulation works to improve galaxy evolution models and quench
star formation rates \citep[e.g.][]{Oppenheimer:2006eq,
    Brook:2012ht, Hopkins:2014dn, Agertz:2014tpa}. However, even if
those simulations are successful in reproducing a set of observations,
it remains unclear and debated whether radiation feedback is
effective, and how it works in detail.

Radiation typically \emph{heats} the gas it interacts with, and though
the heating is relatively gentle compared to AGN and SN feedback, it
may well give an important boost to those other feedback mechanisms
\citep[e.g.][]{Pawlik:2009jv}. \emph{Radiation pressure} may also be
an important feedback mechanism on its own, stirring up the gas in the
inter-stellar medium (ISM) and even generating outflows. Here,
\emph{direct pressure} from ionising radiation can play a role
\citep[e.g.][]{Haehnelt:1995wa, Wise:2012dh, Ceverino:2014dw},
although recent works have relied more on the boost in radiation
pressure that can be gained by reprocessed \emph{multi-scattered}
infrared (IR) radiation, which could in particular be a major feedback
mechanism in optically thick ultra-luminous infrared galaxies, or
ULIRGS \citep[e.g.][]{Murray:2010gh, Thompson:2015fe}. This last
mentioned multi-scattering feedback mechanism in particular has been
under debate in the recent literature. Observationally there is not a
lot of evidence for radiation feedback from star formation, though
recent observations of stellar nurseries hint that its effect on the
ISM is mild and mostly in the form of heating \citep{Lopez:2014cl}. It
is likely though that the nature of the radiation feedback mechanism
depends heavily on the environment, mainly the optical thickness of
the galactic gas.

It does not help that most simulations that invoke some form of
radiation feedback do so with pure hydrodynamics (HD), using subgrid
models and approximations instead of the radiation-hydrodynamics (RHD)
needed to model radiation feedback from first principles.

This is understandable, as radiative transfer is both complex and
costly due to the usually much shorter inherent timescales and large
number of computational dimensions. RHD is still young compared to the
more mature field of HD in galaxy evolution, but in the last decade or
so, increased computational power and the development of new
approaches and algorithms has finally made RHD a feasible prospect in
astronomical and cosmological simulations
\citep[e.g.][]{Petkova:2009fx, Krumholz:2011kl, Pawlik:2011ei,
  Wise:2011iw, Jiang:2012kp, Skinner:2013fca, Norman:2015gw}.

Recently, in \citet[hereafter \RRT{}]{Rosdahl:2013cea}, we presented
an implementation of RHD in the cosmological code \ramses{}
\citep{Teyssier:2002fj}, that we call \ramsesrt{}.  This work focused
on ionising radiation and its interaction with hydrogen and helium via
ionisation heating, which is indeed one of the possibly relevant
physical mechanisms in radiation feedback. However, we still neglected
radiation pressure in that work, which is cited by many of the
aforementioned works as being the main 'culprit' in radiation
feedback.

In this paper, we describe a step towards simulating radiation
feedback in galaxy evolution simulations from first principles, with
the additions to \ramsesrt{} of radiation pressure and reprocessed
dust-coupled multi-scattered radiation. Our new features include a
novel approach to modelling IR radiation \emph{trapping}, that
describes accurately both the optically thin and thick regimes, a
feature that does not come naturally in radiative transfer
implementations, which usually work well in one regime but not the
other. 


This paper is split into two main sections, describing the method
details (\Sec{methods.sec}) and then verification tests
(\Sec{tests.sec}). In the methods section, we begin in
\Sec{rt_eqs.sec} by presenting the basic moment RHD equations to be
solved, focusing on the new aspects of the radiation force and
radiation-dust coupling in the optically thick regime. Then, in
\Sec{solver.sec}, we recall the main ingredients of our existing RHD
solver, and in \Sec{new_solver.sec} we detail the addition of the
radiation pressure and IR-dust interaction. Concluding the methods
section, we present in \Sec{trapping.sec} our innovative approach to
modelling the propagation of IR radiation correctly in \emph{both} the
optically thin and thick limits. The rest of the paper is dedicated to
tests of our implementation, starting with qualitative tests of
radiation field morphology in the optically thin and thick limits
(\Sec{jiang.sec}-\Sec{gonzalez.sec}), going on to test the direct
momentum transfer from photons to gas (\Sec{dir_press_test.sec}), the
correct diffusion of radiation in the optically thick limit
(\Sec{diff_test.sec}-\Sec{3d_diff_test.sec}), and, finally, comparing
our code directly to another RHD implementation in a previously
published experiment of the competition between radiation pressure and
gravity, for which most of our new additions are quite relevant
(\Sec{davistest.sec}).  In the appendix we describe relativistic
corrections to our implementation, the details of which are omitted
from the main text for clarity.

\section{Methods}\label{methods.sec}
RHD has been partially implemented in \ramsesrt{} (\RRT{}), which is
an extension of the adaptive mesh refinement (AMR) code \ramses{}
\citep{Teyssier:2002fj}.  \ramses{} models the interaction of dark
matter, stellar populations and baryonic gas, via gravity, HD and
radiative cooling. The gas evolution is computed using a second-order
Godunov scheme for the Euler equations, while trajectories of
collisionless DM and stellar particles are computed using a
particle-mesh solver.  \ramsesrt{} adds the propagation of photons and
their interaction with gas via photoionisation and heating of hydrogen
and helium. The advection of photons between grid cells is described
with the moment method and the M1 closure relation for the Eddington
tensor. \ramsesrt{} solves the non-equilibrium evolution of the
ionisation fractions of hydrogen and helium, along with ionising
photon fluxes and the temperature in each grid cell.

The goal of the present paper is to extend the RHD implementation in
\ramses{}, adding three important features: i) we now include the
radiative force, which couples the radiation flux to the gas momentum
equation; ii) we introduce a new scheme to recover the proper
asymptotic limit in the radiation diffusion regime, in case the mean
free path is much smaller than the grid spacing; iii) we add
relativistic corrections to the RHD equations, accounting for Doppler
effects up to first order in $v/c$, where $v$ and $c$ are the gas and
light speeds, respectively, and for the work done by the radiation
force on the gas.  In this section, we will review the main
characteristics of the \ramsesrt{} solver before discussing our new
numerical scheme for the radiation force and for the preservation of
the asymptotic diffusion regime. We will omit the order $v/c$
relativistic corrections, which will be described in more detail in
the Appendix.

\subsection{The RHD Equations}\label{rt_eqs.sec}
We describe here the moment equations solved in \ramsesrt{}, outlining
the role played by the radiation force.  

As detailed in \RRT{}, we use an important approximation to speed up
our explicit scheme for radiation advection, where the time-step
scales inversely with the speed of light $c$. In this so-called
reduced speed of light approximation, we simply decrease the speed of
light, typically by $1-3$ orders of magnitude\footnote{This
  approximation is valid only if the modified light crossing time is
  still short compared to the sound crossing time, the recombination
  time, and the advection time in the flow.  If this is not the case,
  then the reduced speed of light approximation is invalid and one has
  to rely on either RT subcycles \citep{Aubert:2008jj} or implicit
  time integration \citep{Commercon:2014ca}.}. In this paper, we thus
make an important distinction between $c$, the actual speed of light,
and $\cred$, the reduced speed of light.

The starting point in deriving the RHD equations is the radiation
specific intensity $I_{\nu}({\bf x},{\bf n},t)$, describing the
radiation flow (CGS units of $\eflux \, \hz^{-1} \,
\rm{rad}^{-2}$)\footnote{We will use CGS units
  (centimeters-grams-seconds) to clarify variable dimensions, but
  these are obviously interchangeable for other units systems.}, over
the dimensions of frequency $\nu$, location ${\bf x}$, unit direction
${\bf n}$, and time $t$. The evolution of the specific intensity is
described by the radiative transfer (RT) equation:
\begin{align}\label{RT.eq}
\frac{1}{\cred}\frac{\partial I_\nu}{\partial t} + 
{\bf n}\cdot \nabla I_\nu = \eta_\nu - \kappa_\nu \rho I_\nu,
\end{align}
where $\kappa_\nu$ is the gas opacity, ($\ccg$), $\rho$ the gas
density ($\gcc$), and $\eta_\nu$ the plasma emissivity ($\ergs \, \cci
\, \hz^{-1} \, \rm{rad}^{-2}$, usually assumed to be isotropic).
 
We define the radiation energy density $E$ ($\edens$), the radiation
flux ${\bf F}$ ($\eflux$), and the radiation pressure ${\mathbb P}$
($\edens$), {\it in a group of photons} over a specified frequency
range, as \emph{moments} (i.e. averages) of the radiation intensity
over solid angle $\Omega$ and frequency:
\begin{align}
  E({\bf x},t)&=\frac{1}{\cred}\int_\nu \int_{4\pi} I_\nu ({\bf x},{\bf
    n},t) ~d\nu \, d{\bf \Omega}, \label{mom1.eq}\\ 
{\bf F}({\bf x},t)&=\int_\nu \int_{4\pi} I_\nu ({\bf x},{\bf n},t) 
    ~{\bf n} ~d\nu \, d{\bf \Omega}, \label{mom2.eq}\\
{\mathbb P}({\bf x},t)&=\frac{1}{\cred}\int_\nu \int_{4\pi} I_\nu 
    ({\bf x},{\bf n},t)  {\bf n} \otimes  {\bf n}  ~d\nu \, d{\bf
    \Omega}, \label{mom3.eq}
\end{align}
where $\otimes$ denotes the outer product.  Taking the zeroth and
first moments of \Eeq{RT.eq} and substituting the definitions
(\ref{mom1.eq}-\ref{mom3.eq}) yields the well-known moment equations
of radiation energy and flux \citep[e.g.][]{Mihalas:1984vm}:
\begin{align}
  \frac{\partial E}{\partial t} + \nabla \cdot {\bf F} &= S -
  \kE \rho \cred E, \label{momeq1.eq}\\
  \frac{1}{\cred}\frac{\partial {\bf F}}{\partial t} + \cred \nabla
  \cdot {\mathbb P} &= - \kF \rho {\bf F}, \label{momeq2.eq}
\end{align}
where $\kE$ and $\kF$ are respectively the radiation energy and flux
weighted mean opacities, and the source function $S$ is the integral
of the emissivity over all solid angles and over the photon groups
frequency range. With multiple photon groups, a separate set of moment
equations exists for each group, which should in principle be denoted
by photon group subscripts, i.e. $E_i$, ${\bf F}_i$, ${\mathbb P}_i$,
$S_i$, $\kappa_{E,i}$, and $\kappa_{F,i}$. For the sake of simplicity,
we omit those subscripts, unless they are required for clarification.

If the system under study is close to Local Thermodynamical
Equilibrium (LTE), where the gas emits as a blackbody, and the photon
group covers a sufficiently large frequency range, the source function
can be approximated by the frequency integral of a Planckian,
\begin{align}
  S = \kP \rho c a T^4,
\end{align}
where $a$ is the radiation constant, $\kP$ is the Planck mean opacity,
and $T$ is the gas temperature.  This approximation is often used to
describe the coupling between dust and IR radiation in the ISM
\cite[][chapter 6]{Mihalas:1984vm}. We assume a \emph{single-fluid}
system in this work, where the gas and dust are also in LTE, i.e. at
the same temperature.
Note that in the previous equations, the opacities are computed in the
{\it comoving} frame, moving with the gas, while the radiation moments
are defined in the laboratory (or lab) frame. We ignore Doppler
effects of these relative motions in the main text. However including
them for non-relativistic flows introduces important additional terms
which are described in the Appendix.

If one assumes that the spectral energy distribution is close to a
Planckian, then $\kE=\kP$.  Another traditional approximation, when
the fluid-radiation system is close to LTE and the optical depth is
large, is to take $\kF \simeq \kR$, where the latter is the
Rosseland mean. Under these approximations, valid only for systems
close to LTE (such as for ISM dust and IR radiation), equations
(\ref{momeq1.eq}-\ref{momeq2.eq}) simplify into
\begin{align}
\frac{\partial E}{\partial t} + \nabla \cdot {\bf F} 
&= \kP \rho \left(c a T^4 - \cred E \right), \label{mom_E.eq} \\
\frac{\partial {\bf F}}{\partial t} 
+ \cred^2 \nabla \cdot {\mathbb P} &= - \kR \rho \cred {\bf F}. 
\label{mom_F.eq}
\end{align}
These equations are \emph{not valid} in the optically thin regime and
for systems far from LTE, such as for ionising radiation coupled to
the non-equilibrium chemistry of hydrogen and helium.  Under such
conditions, one can instead use a template spectrum, usually the
Spectral Energy Distribution (SED) of stellar populations, to compute
the average dust opacities (see \RRT{}).

The HD equations must be modified to account for the transfer of
energy and momentum between radiation and gas. The fluid energy
equation describes the evolution of the gas energy density
\begin{align}
  E_{\rm gas} = \frac{1}{2} \, \rho v^2 + e,
\end{align}
where the right hand side (RHS) terms are kinetic energy, with $v$ the
gas speed, and internal or `thermal' energy $e$. Assuming LTE, the
fluid energy equation becomes
\begin{align}
  \frac{\partial E_{\rm gas}}{\partial t} 
  + \nabla \cdot \left( {\bf v} (E_{\rm gas}+P) \right) 
  = \rho {\bf g} \cdot {\bf v} \label{lte_egy.eq}
  + \Lambda + \kP \rho \left(\cred E - c a T^4 \right),
\end{align}
where ${\bf v}$ and $P$ are the gas velocity and pressure, ${\bf g}$
is the local gravitational acceleration, and $\Lambda$ represents
cooling/heating via thermochemical processes (see \RRT{}). The new
term here is the last one on the RHS, describing the internal energy
exchange between the gas and the radiation field.

The fluid momentum equation becomes
\begin{align} \label{fluid_mom.eq}
  \frac{\partial \rho {\bf v}}{\partial t} 
  + \nabla \cdot \left( \rho {\bf v} \otimes {\bf v}+P {\mathbb I}
  \right) = \rho {\bf g} + \frac{\kR \rho}{c} {\bf F}, 
\end{align}
where ${\mathbb I}$ is the identity matrix.  Here the new term is
again the last one on the RHS, describing the radiation momentum
absorbed by the gas.  Note that the work done by the radiation force
is absent. These terms of order $v/c$ are introduced in the Appendix
as a relativistic correction, but we omit them from the main text for
the sake of simplicity.

\subsection{The Radiation Solver}\label{solver.sec}
\ramsesrt{} solves the radiation advection equations
(\ref{mom_E.eq}-\ref{mom_F.eq}) using the M1 closure for the Eddington
tensor, first introduced by \cite{Levermore:1984cs}.  In this
approximation, the Eddington tensor, defined as ${\mathbb P}={\mathbb
  D}E$, is given explicitly by a simple {\it local} relation
\begin{align}\label{RTmom1a.eq}
{\mathbb D} = \frac{1-\chi}{2} {\mathbb I} 
+ \frac{3\chi-1}{2}~{\bf n}\otimes {\bf n},
\end{align}
where ${\bf n}={\bf F}/|{\bf F}|$ and $\chi$ depends only on the
reduced flux,
\begin{align}\label{redFlux.eq}
  f = \frac{\left| {\bf F}\right|}{\cred E},
\end{align}
as
\begin{align}\label{RTmom1b.eq}
  \chi(f) = \frac{3+4f^2}{5+2\sqrt{4-3f^2}}.
\end{align}
It is based on the assumption that the angular distribution of the
radiation intensity can be approximated by a Lorentz-boosted
Planckian, in the direction of the radiation flux.  This approximation
recovers the asymptotic limit of the diffusion regime, when $f \ll 1$,
so that $\chi \simeq 1/3$ and ${\mathbb D} \simeq {\mathbb I}/3$.  It
also describes well the free streaming of radiation from a single
source, when $f \simeq 1$, so that $\chi \simeq 1$ and ${\mathbb D}
\simeq {\bf n} \otimes {\bf n}$.  In the intermediate regime, or in
the presence of multiple sources, this is only an approximation, and
the model must therefore be compared to existing exact solutions to
assess its range of validity \citep[][\RRT{}]{Aubert:2008jj}.

A very important consequence of the M1 closure is that the resulting
system of conservation laws (ignoring the source terms) is hyperbolic,
and can therefore be integrated numerically using a classical Godunov
scheme \citep{Aubert:2008jj}, and an operator split approach, where
the radiation variables $E$ and ${\bf F}$ in each cell are modified
first using a conservative and {\it explicit} update from their
intercell fluxes, and the source terms are included in a second step
using a local, {\it implicit}, sub-cycling thermochemistry module
\citep[][\RRT{}]{Aubert:2008jj}.

Stability of the numerical integration for the transport step is
ensured using proper upwinding to compute the numerical flux, using a
{\it Riemann solver}.  In this paper, we use the Global Lax Friedrich
(GLF) Riemann solver\footnote{\ramsesrt{} also offers the possibility
  to use the Harten-–Lax-–van Leer (HLL) intercell flux function,
  which is less diffusive than GLF, but also produces less spherically
  symmetric radiation from stars, as we showed in \RRT{}. Our method
  for radiation trapping in the optically thick limit, which we
  develop in this paper, is however only strictly compatible with GLF,
  so we do not include the HLL function in the current work. Since we
  prefer the GLF function over HLL, which produces asymmetric
  radiation around stellar sources, we do not have immediate plans to
  adopt radiation trapping for HLL}.
\citep[see][\RRT{}]{Aubert:2008jj}, for which the interface radiation
flux is explicitly
\begin{align} \label{GLF.eq}
  \bF_{1/2}\left( \state_{\rm L}, \state_{\rm R} \right) 
  = \frac{\bF_{\rm R}+\bF_{\rm L}}{2} 
  - \frac{\cred}{2}\left( E_{\rm R} - E_{\rm L} \right),
\end{align}
where $\state=\left( \bF, E \right)$ is a cell state, the `1/2'
subscript refers to the Godunov intercell state, that we use to
perform the final conservative update of the radiation energy, and the
subscripts `L' and `R' refer to the neighbouring left and right
cells. A similar formula holds for the intercell Eddington tensor to
conservatively update the radiation flux.  The first term on the RHS
of \Eeq{GLF.eq} is the average of the right and left cells radiation
fluxes. This term alone would give a second-order but {\it unstable}
solution. The second term on the right-hand side of \Eeq{GLF.eq} is
proportional to the difference of the right and left cell radiation
densities. This is the stabilising term, also called the numerical
diffusion term. Indeed, one can formally rewrite the numerical flux as
\begin{align}\label{GLF_formal.eq}
  \bF_{1/2} = \frac{\bF_{\rm R}+\bF_{\rm L}}{2} 
  - \frac{\cred \Delta x}{2} \frac{\partial E}{\partial x},
\end{align}
where $\Delta x$ is the width of the or cell. We now see explicitly
the numerical diffusion coefficient as $\nu_{\rm num}= \cred \Delta x
/2$.  We will use these numerical concepts in Section
\ref{trapping.sec}.

\subsection{A New RHD Solver}\label{new_solver.sec}
The microscopic processes that are already included in \ramsesrt{}
(see \RRT{}) are the non-equilibrium chemistry of hydrogen and helium
coupled to the ionising radiation. We now describe the new features in
\ramsesrt{} which can be used to model the coupling between dust and
IR radiation, and to model the injection of momentum into the gas by
the radiation flux.

\subsubsection{Modified moment RT equations, for IR and higher energy
  photons}
In \ramsesrt{}, we now make a distinction between the group of IR
photons and all other, higher-energy, groups. The IR photons are
assumed to cover the energy range of dust emission and to be in LTE
with the dust particles, exchanging energy via absorption and
re-emission. Other groups, however, span energies above the dust
emission. These photons can be absorbed by the dust, as well as by
hydrogen and helium via photoionisation, but the dust-absorbed energy
is re-emitted at lower (IR) energies. Thus, the IR photons can be seen
as `multi-scattered', while all other photons are `single scattered'.

For a group $i\ne$IR of non-IR photons, the moment RT equations,
following from Eqs. (\ref{momeq1.eq}-\ref{momeq2.eq}), are unchanged from
what we presented in \RRT{}, save for new dust absorption terms:
\begin{align}
  \frac{\partial E_i}{\partial t} + \nabla \cdot {\bf F_i} =
  -\sum_j^{\hi,\hei,\heii} n_j \csn_{ij} \cred E_i
  + \dot{E}_i 
  - \ki \rho \cred E_i, \label{RTmom1.eq}\\
  \frac{\partial {\bf F_i}}{\partial t} + \cred^2 \nabla \cdot
  {\mathbb P_i} = -\sum_j^{\hi,\hei,\heii} n_j \csn_{ij} \cred {\bf
    F}_i  - \ki \rho \cred {\bf F_i}. \label{RTmom2.eq}
\end{align}
Here we sum over the hydrogen and helium species $j$ which absorb
ionising photons, with $\csn_{ij}$ denoting the ionisation cross
section ($\cs$) between photon group $i$ and ion species $j$, which is
zero for non-ionising photons. $\dot{E}$ is the rate of emission from
point sources (stars, AGN) and hydrogen/helium recombinations. The
last terms in each equation represent dust absorption, which scales
with the dust-opacity ($\ki$) and the gas density.

The dust absorbed energy is re-emitted into the IR photon group, for
which the RT equations are
\begin{align}
  \frac{\partial E_{\rm IR}}{\partial t} + \nabla \cdot {\bf F_{\rm
      IR}} &= \ \kP \rho \left( c a T^4 - \cred E_{\rm IR}\right) +
  \dot{E}_{\rm IR}  \nonumber \\
  &+\sum_i^{\rm other \ groups} \ki \rho \cred E_i, \label{momeq_IR_E.eq} \\
  \frac{\partial {\bf F_{\rm IR}}}{\partial t} + \cred^2 \nabla \cdot
  {\mathbb P_{\rm IR}} &= - \kR \rho \cred {\bf F_{\rm
      IR}}. \label{momeq_IR_F.eq}
\end{align}
These equations are the same as the previous equations
(\ref{RTmom1.eq}-\ref{RTmom2.eq}) for non-IR photons, except that
i) we omit photoionisation/recombination terms (in
$\dot{E}_{\rm IR}$), as these photons have sub-ionising energies, ii)
the negative dust absorption terms in the previous equations become
additive terms here, representing dust re-emission into the IR group,
and iii) we have added the first RHS term, which describes the
coupling between IR radiation density and the gas (dust) temperature.

A great deal of complex physics is encapsulated inside $\ki$, $\kR$,
and $\kP$, which depend on temperature, the dust content, and the
exact shape of the radiation spectrum. One can use existing models for
temperature-dependent dust opacities \citep[e.g.][]{Draine:2007kk},
assume that the dust content scales with metallicity, and include a
cutoff at $T \ga 1000$ K to model dust sublimation. In this work,
however, we consider only constant values for the photon opacities,
except for Section \ref{davistest.sec}, where we use simple
temperature dependent functions. Updating the opacities to more
complex forms is a straightforward addition to the code, and often
specific to the problem at hand and the level of detail one seeks to
achieve. We defer those considerations to future works.

As described in detail in \RRT{}, the RT moment equations are solved,
after the HD step, with an operator splitting approach, where we solve
in sequence the advection terms and the source/sink terms over an RHD
time-step, for all cells in a given refinement level. The advection is
solved explicitly and the source/sink terms are solved
quasi-implicitly, together with the gas temperature, using
thermochemistry sub-cycling. The only non-trivial addition to the
solver is the coupling term for the gas and radiation, i.e. the first
term on the RHS of \Eeq{momeq_IR_E.eq}, which is described next.

\subsubsection{IR-dust temperature coupling}
Ignoring advection terms and other sources of photon
absorption/emission and gas cooling/heating, which are described in
\RRT{}, the coupling between the IR energy density, $E_{\rm IR}$, and
the gas internal energy density, $e$, follows from
Eqs. (\ref{mom_E.eq}) and (\ref{lte_egy.eq}), respectively:
\begin{align}
  \frac{\partial E_{\rm IR}}{\partial t}
  &= \kP \rho \left( c a T^4 - \cred E_{\rm IR}\right), \label{TEcpl1.eq}\\
  \frac{\partial e}{\partial t} &= \kP \rho \left(\cred E_{\rm IR}
    - c a T^4 \right) \label{TEcpl2.eq}.
\end{align}
These equations are solved in each thermochemistry substep
\emph{after} the updates of radiation energy density and gas
temperature via other terms of absorption, emission, heating, and
cooling. Keeping in mind the strong coupling between radiation and
temperature, we solve semi-implicitly using a linear approach. In this
formulation, the change in the state vector $\state_E \equiv \left(
  E_{\rm IR}, e \right)$, over the thermochemistry time-step of
length $\Delta t$, is
\begin{align}
\Delta \state_E = \dot{\state}_E \Delta t \ 
   \left( {\mathbb I} - \Jacobian \Delta t \right)^{-1}, 
\end{align}
where $\dot{\state}_E$ is the RHS of
Eqs. (\ref{TEcpl1.eq}-\ref{TEcpl2.eq}), and $\Jacobian =
\frac{\partial \dot{\state}_E}{\partial \state_E}$ is the Jacobian
matrix, each evaluated at the start of $\Delta t$.

Taking advantage of the symmetry of the problem ($\Delta E_{\rm
  IR}=-\Delta e$), the update over $\Delta t$ is obtained by
\begin{align}\label{TEcpl3.eq}
  \Delta E_{\rm IR}=-\Delta e = \frac{caT^4-\cred E_{\rm
      IR}}{\left( \kP \rho \Delta t \right)^{-1} + \cred + 4 caT^3
    C_{\rm V}^{-1}},
\end{align}
where $C_{\rm V}=\left( \frac{\partial e}{\partial T }\right)_{\rm
  V} = \frac{\rho \kb}{ \mp \mu (\gamma-1)}$ is the heat capacity at
constant volume, $\kb$ the Boltzmann constant, $\mu$ the average
particle mass in units of the proton mass $\mp$, and $\gamma$ is the
ratio of specific heats.
 
After the update of temperature and IR energy via \Eeq{TEcpl3.eq}, we
re-apply the $10\%$ thermochemistry rule (\RRT{}): if either $T$ or
$E_{\rm IR}$ (or both) was changed by more than $10\%$ from the
original value, the entire thermochemistry sub-step is repeated with
half the time-step length.

\subsubsection{Momentum transfer from photons to gas}
In the framework of the RHD method, the fluid momentum equation is
\begin{align}
  \frac{\partial \rho {\bf v}}{\partial t} 
  + \nabla \cdot \left( \rho {\bf v} \otimes {\bf v}+P {\mathbb I}
  \right) = \rho {\bf g} + \momRadRate.
\end{align}
This is the same as \Eeq{fluid_mom.eq}, but generalised to the total
local momentum absorption rate, per unit volume, from all photon
groups via all radiation interactions (not only radiation-dust
interactions):
\begin{align}
  \momRadRate = 
  \sum_i^{\rm groups} 
  \frac{\bF_i}{c} \left(  \ki \rho + 
    \sum_j^{\hi,\hei,\heii}  \csn_{ij} n_j \right).
\end{align}

The momentum transfer is implemented with an operator split approach,
adding to the gas momentum in each RHD step after the thermochemistry
step.  Since both photon fluxes and absorber densities may change
substantially during the sub-cycling of the thermochemistry equations
over a single RHD time-step, $\dtrhd$, we collect the absorbed
momentum density over the subcycles, whose sub-time-steps are limited
such as to change the evolved quantities only by a small fraction
($10\%$) per substep:
\begin{align}
  \Delta {\bf{p}_{\gamma}} =  \sum_k \Delta t_k
  \sum_i^{\rm groups} \frac{\bF_{i,k+1}}{c} \left(  \ki \rho +
    \sum_j^{\hi,\hei,\heii} 
     \csn_{ij} n_{j,k+1} \right) \label{dir_mom.eq}.
\end{align}
Here the outermost sum is over the thermochemistry sub-steps (with
$\sum_k \Delta t_k = \dtrhd$).  At the end of the thermochemistry
sub-cycling of a cell, the total absorbed photon momentum density
vector $\Delta {\bf{p}_{\gamma}}$ is added to the gas momentum, and
the gas specific total energy is updated to reflect the change in
kinetic energy.

In addition to the direct radiation pressure just described, radiation
pressure from isotropic diffusive IR radiation is also implemented in
\ramsesrt{}, as we will discuss in the next subsection.

\subsection{Preserving the Asymptotic Diffusion Limit}\label{trapping.sec}
The \emph{diffusion limit} is reached when the optical depth of the
LTE radiation becomes unresolved and the photons propagate in a random
walk\footnote{This section concerns only the IR photon group, since
  other groups are assumed to be single scattering.}. Then, since $F
\ll \cred E$, we get for the Eddington tensor (\Eq{RTmom1a.eq})
${\mathbb D} = {\mathbb I}/3$. In this case, we reach the asymptotic
regime where \Eeq{mom_F.eq} reduces to a static form
\citep[see][Sec. 80]{Mihalas:1984vm}\footnote{The ratio between the
  time-dependent and static flux terms in \Eeq{mom_F.eq} is
\begin{align}
  \frac {\frac{\partial {\bf F}}{\partial t}}
  {\kR \rho \cred {\bf F}}
  \sim
  \frac{\lR}{\cred \dt} 
  = \left( \frac{\lR}{\dx} \right)^2,
\end{align}
where we use the fact that a traveled distance $\dx$ requires
$\left(\dx/\lR \right)^2$ interactions in a random walk, and hence
the time to travel this distance is $\dt=\frac{\dx^2}{\cred \lR}$. If
$\lR \ll \dx$, the time-dependent flux term is thus negligible, and we
can use the static diffusion form (\Eq{Flux_asymp.eq}).  }, giving
\begin{align} \label{Flux_asymp.eq} 
{\bf F} \simeq - \frac{\cred \lR}{3}\nabla E,
\end{align}
\noindent
where $\lR=\left(\kR \rho \right)^{-1}$ is the mean free
path.  This equation expresses the fact that in this regime, radiation
is a diffusive process, with diffusion coefficient $\nu_{\rm rad}=
\cred \lR / 3$.  The previous derivation for our numerical
scheme (see \Eq{GLF.eq}) explicitly demonstrates that in the diffusion
limit, the numerical diffusion of our M1 solver dominates over the
true radiation diffusion when
\begin{align}\label{diffusion_criterion.eq}
  \nu_{\rm num} > \nu_{\rm rad}~~~{\rm or}~~~3 \Delta x > 2 \lambda_{\rm
    R}. 
\end{align}
\noindent 
This last inequality is likely to occur in optically thick regions,
where the optical depth of the cell, $\tauc = \Delta x / \lambda_{\rm
  R}$, is larger than 1.

As discussed in \cite{Liu:1987ga} and \cite{Bouchut:2004wk}, if the
\Eeq{diffusion_criterion.eq} inequality occurs, operator splitting is
not valid anymore, as source terms become stiff compared to the
hyperbolic transport terms. The numerical result becomes severely
inaccurate: radiation propagates with an effective mean-free-path
equal to the cell size, much larger than the true mean-free-path,
manifesting in photons which travel much too fast through the volume,
compared to \Eeq{Flux_asymp.eq}.

One possibility to resolve the problem and recover the correct
diffusion of photons is to exploit the AMR technique and refine the
grid adaptively so that $\Delta x$ always stays smaller than, say,
$\lambda_{\rm R}/4$. This is unfortunately not always possible in
realistic astrophysical applications where the opacity can be a highly
non-linear function of temperature and density.

We now propose two different techniques to modify our base scheme in
order to preserve the asymptotic diffusion regime posed by
\Eeq{Flux_asymp.eq}: i) a modification of the Godunov flux that takes
into account the diffusion source term (\Sec{godflux.sec}), and ii)
the addition of a new photon (sub-) group that we call {\it trapped
  photons} (\Sec{trapping2.sec}). As opposed to {\it streaming
  photons}, these new photons are strictly isotropic in angular space.

\subsubsection{Asymptote-preserving Godunov fluxes}\label{godflux.sec}
Following the methodology presented in \cite{Berthon:2007fh}, it is
possible to correct for the effect of radiation diffusion by
explicitly taking into account the source terms in the Riemann
solver. The Riemann solution becomes much more complicated
\citep[see][]{Berthon:2007fh}, but can be approximated by a simple
modification of the intercell flux (\Eq{GLF.eq}) as
\begin{align} \label{modifGLF}
  \bF_{1/2} = \bF_{1/2}(\alpha_{\rm L} \state_{\rm L}, \alpha_{\rm R}
  \state_{\rm R}),
\end{align}
\noindent 
where \cite{Berthon:2007fh} introduced the new function
$\alpha(\tauc)$, which is, in case one uses the GLF numerical flux,
\begin{align}
\alpha(\tauc)=\frac{1}{1+\frac{3}{2}\tauc}.
\end{align}
This function encodes the modification to the Riemann solver that
accounts for the source terms. It satisfies
\begin{align}
\alpha \rightarrow 1 &~~{\rm when}~~ \tauc  \rightarrow 0,
~~~~~~~~{\rm and}  \nonumber \\
\alpha \rightarrow \frac{2}{3\tauc} &~~{\rm when}~~ \tauc \rightarrow
+\infty. \nonumber
\end{align}

Our goal is to recover the correct asymptotic limit in the optically
thick regime.  Using \Eeq{GLF_formal.eq} with the above modification,
we indeed find, assuming for simplicity that the mean free path is
uniform, that the numerical flux has the correct asymptotic behaviour
given by \Eeq{Flux_asymp.eq}:
\begin{align} \label{GLF_mod1.eq}
  \bF_{1/2} &\simeq \frac{2\lambda_{\rm R}}{3\Delta x} 
  \frac{\left( \bF_{\rm R} + \bF_{\rm L}\right)}{2} 
  - \frac{\cred \lambda_{\rm R}}{3} 
  \frac{\left( E_{\rm R} - E_{\rm L}\right)}{\Delta x} \nonumber \\
  &\simeq - \frac{\cred \lambda_{\rm R}}{3} 
  \frac{\left( E_{\rm R} - E_{\rm L}\right)}{\Delta x}.
\end{align}
The latter equality comes from the fact that in the limit of optically
thick cells, the absorption terms in \Eeq{momeq_IR_E.eq} naturally lead
to $F \ll \cred E$.

\subsubsection{Trapped versus streaming photons}\label{trapping2.sec}

Although the previous method allows us to upgrade, in a
straightforward way, our M1 hyperbolic solver for the transport of
radiation in a dense, optically thick medium, we have instead
implemented in \ramsesrt{} an alternative technique, that turns out to
be equivalent to the previous one, but allows for a more accurate
treatment of the diffusion limit, where trapped photons are advected
with the gas, and radiation pressure, along with the work performed by
that pressure, is naturally accounted for.

Our technique is based on the "IDSA methodology" (Isotropic Diffusion
Source Approximation), proposed by \cite{Liebendorfer:2009kw} in the
context of neutrino transport in core collapse supernovae. The idea is
to introduce two different IR photon groups spanning the same
frequency range, splitting the total IR radiation energy into a {\it
  trapped} radiation energy variable $E_t$ and a {\it streaming}
radiation energy variable $E_s$ satisfying $E=E_t+E_s$. The difference
between the trapped and streaming photons is that the former are
assumed to be strictly isotropic in angular space. They correspond to
the asymptotic limit of vanishingly small mean free path, for which
the radiation flux is strictly zero. We can then rewrite the radiation
moment equations, (\ref{momeq_IR_E.eq}-\ref{momeq_IR_F.eq}), using
${\bf F_t}=0$ as
\begin{align}
\frac{\partial E_t}{\partial t} + 
   \frac{\partial E_s}{\partial t} + \nabla \cdot {\bf F_s}  &= \kP \rho
   \left(c a T^4  - \cred E_t - \cred E_s \right) + \dot{E}, 
   \label{Etstot.eq}\\
\frac{\partial {\bf F_s}}{\partial t} 
   + \frac{\cred^2}{3} \nabla E_t + \cred^2 \nabla \cdot  {\mathbb P}_s 
    &= - \kR \rho \cred {\bf F}_s, \label{t-s-flux.eq}
\end{align}
where we used the fact that ${\mathbb P}_t= E_t \, {\mathbb I}/3$
(\Eq{RTmom1a.eq}) since trapped photons are isotropic, and we enclosed
the isotropic emission terms from gas, stars, AGN, and other photon
groups in \Eeq{momeq_IR_F.eq} under one term, $\dot{E}$.

\cite{Liebendorfer:2009kw} proposed to split the previous system into
two sets of equations, one describing the trapped photons only,
\begin{align} \label{Et_adv.eq}
  \frac{\partial E_t}{\partial t}  
  = \kP \rho \left(c a T^4  - \cred E_t \right) + \dot{E},
\end{align}
where the isotropic source of radiation is assigned naturally to the
trapped component, and a second one describing the streaming photons
only, with
\begin{align}
\frac{\partial E_s}{\partial t} +  \nabla \cdot {\bf F_s} 
   &= - \kP \rho \cred  E_s, \\
\frac{\partial {\bf F_s}}{\partial t} 
   + \cred^2 \nabla \cdot  {\mathbb P}_s 
   &= - \kR \rho \cred {\bf F}_s - \frac{\cred^2}{3} \nabla E_t, 
   \label{Lieb2.eq}
\end{align}
\noindent
where the last two equations are our standard moment equations
(\ref{momeq_IR_E.eq}-\ref{momeq_IR_F.eq}), only with modified source
terms. This is the system that we would like to solve using our
Godunov scheme. In the \cite{Liebendorfer:2009kw} approach, the next
step is to introduce an additional fictitious source term describing
the energy exchange between trapped and streaming photons (noted
$\Sigma$ in the IDSA methodology).

We follow a different route, analysing the asymptotic diffusion
regime, which gives a straightforward decomposition between trapped
and streaming photons. Indeed, in the diffusion limit, we have $E_s
\ll E_t$, and \Eeq{Lieb2.eq} becomes
\begin{align} \label{Fstream.eq}
{\bf F}_s \simeq - \frac{\cred \lambda_R}{3} \nabla E_t.
\end{align}
On the other hand, we know that the numerical diffusion term for
streaming photons in the GLF flux function of our Godunov scheme
(\Eq{GLF_formal.eq}) is
\begin{align}
{\bf F}_s \simeq - \frac{\cred \Delta x}{2} \nabla E_s.
\end{align}
It is then straightforward to make a partition between streaming and
trapped photons, such that \Eeq{Fstream.eq} is correctly retrieved in
our photon advection scheme. The relations which ensure this are
\begin{align}\label{trappedTot.eq}
  E_t = \frac{3\tauc}{2} E_s~~~{\rm and}~~~E=E_t+E_s,
\end{align}
i.e.
\begin{align}
  E_s=\frac{2}{2+3\tauc}E{\rm ,}~~~{\bf F_s}={\bf F},
  \label{E_stream_org.eq} \\
  E_t=\frac{3\tauc}{2+3\tauc}E{\rm ,}~~~{\bf F_t}=0. 
  \label{E_trap_org.eq}
\end{align}
Using this partition, we can describe our streaming photon group
with the classical Godunov solver (\Eq{GLF.eq}) without the additional
source term in \Eeq{Lieb2.eq}, namely
\begin{align}
\frac{\partial {\bf F_s}}{\partial t} 
   + \cred^2 \nabla \cdot  {\mathbb P}_s = - \kR \rho \cred {\bf F}_s,
\end{align}
and still get the correct asymptotic diffusion limit of the mixed
trapped/streaming system.

In other words, by making the partition of
Eqs. (\ref{E_stream_org.eq}-\ref{E_trap_org.eq}) between streaming and
trapped photons, in all cells, \emph{before} each photon advection
step, the streaming photon variables, $E_s$ and $\bF_s$, can be
advected using Eqs. (\ref{momeq_IR_E.eq}-\ref{momeq_IR_F.eq}), without
any modification to the RT advection solver. The RT solver, however,
does not touch the trapped photon variable, $E_t$. We
\emph{de-partition} between the trapped and streaming photons before
the thermochemistry step, such that thermochemistry is performed on
the total photon density and flux, and \emph{re-partition} once the
thermochemistry step is finished, such that the advection is correctly
performed in the diffusion limit. The modification to the RHD code to
correctly account for the diffusion limit is thus limited to a single
new variable ($E_t$), and a few lines of code before and after the
call to the thermochemistry.

In addition to this simple modification, we need to also make sure
that i) the trapped photons are advected with the gas, ii) that
radiation pressure from the trapped photons is correctly accounted
for, and iii) that the $PdV$ work done on the gas by the trapped
radiation pressure is accounted for, by reducing the trapped radiation
energy accordingly. Fortunately, all these features are automatically
acquired in \ramses{}, by storing the trapped radiation as a
non-thermal energy variable. Non-thermal energy variables are a new
feature in \ramses{}, adding up the total energy density and pressure
which is used in the classical Euler HD equations \citep[see
e.g.][Eqs. 39-40]{Rosdahl:2013cea}, and they behave just like the
thermal energy. In other words, the trapped radiation energy is
correctly advected with the gas, the trapped radiation pressure is
correctly accounted for, and so is the $PdV$ work done by the trapped
radiation. These relativistic details are covered in
\App{vc2.app}. The equation of state relating the trapped radiation
energy and pressure, is
\begin{align} \label{Pnt.eq}
  \Prad= \frac{\cred}{c} \frac{E_t}{3}.
\end{align}
The radiative force is computed as the sum of the trapped and
streaming contributions (from \Eq{t-s-flux.eq}), which, in our model,
is also equivalent to the Godunov GLF flux of the streaming
photons. The fluid momentum equation (\ref{fluid_mom.eq}) thus becomes
\begin{align}\label{mom_trapped.eq}
  \frac{\partial \rho {\bf v}}{\partial t} 
  + \nabla \cdot \left( \rho {\bf v} \otimes {\bf v}
  +(P+\Prad) {\mathbb I} \right) =\frac{\kR \rho}{c}{{\bf F}_s} 
  + \rho {\bf g},
\end{align} 
where we omit the contributions from single scattering photon groups,
which have the same form as the first term on the RHS.  In the
diffusion limit, for which $E_s \ll E_t$ and $\bF_s\approx 0$, we
recover the regime where the radiative force is equal to the radiative
pressure gradient
\begin{align}
  \frac{\partial \rho {\bf v}}{\partial t} + \nabla \cdot \left( \rho
    {\bf v} \otimes {\bf v} + P  {\mathbb I} \right) = -
  \frac{\cred}{3 c}\nabla \Etrap +\rho {\bf g}.
\end{align} 

With the partition given by
Eqs. (\ref{E_stream_org.eq}-\ref{E_trap_org.eq}), trapped photons are
only generated in regions of the flow where the mean free path is
smaller than the cell size. In opposite situations where the mean free
path is large enough, it is desirable to make sure that the fraction
of trapped photons very quickly converges to zero. We therefore modify
our trapped versus streaming photons distribution using
\begin{align}
E_s & =\left[1-\exp{\left(-\frac{2}{3\tauc}\right)}\right] \ E{\rm ,} 
   \label{E_stream_mod.eq} \\
E_t & =\exp{\left( -\frac{2}{3\tauc} \right)}\ E. \label{E_trap_mod.eq}
\end{align}
This model has the same optically thick limit as the original one,
(Eqs. \ref{E_stream_org.eq}-\ref{E_trap_org.eq}) but trapped photons vanish
much faster in the optically thin limit.

To summarise, our new method starts by initialising the trapped and
streaming radiation variables using
Eqs. (\ref{E_stream_mod.eq}-\ref{E_trap_mod.eq}). Only the streaming
photons are advected using our original Godunov scheme,
\begin{align}
\frac{\partial E_s}{\partial t} +  \nabla \cdot {\bf F_s} 
= - \kP \rho \cred  E_s, \label{Es_adv.eq} \\
\frac{\partial {\bf F_s}}{\partial t} 
+ \cred^2 \nabla \cdot  {\mathbb P}_s = - \kR \rho \cred {\bf F}_s.
\end{align}
For the thermochemistry, including the radiation/matter coupling term,
the IR radiation used is the sum of the free streaming and trapped
photons,
\begin{align}
E=E_t+E_s{\rm ,}~~~{\bf F}={\bf F_s}.
\end{align}
In our operator splitting approach, the streaming radiation density is
in practice advected with \Eeq{Es_adv.eq} with the ${\rm RHS}=0$,
while the RHSs of Eqs. (\ref{Et_adv.eq}) and (\ref{Es_adv.eq}) are
accounted for in the thermochemical coupling of the dust temperature
to the \emph{total} IR radiation temperature, as in
Eqs. (\ref{TEcpl1.eq}-\ref{TEcpl2.eq}):
\begin{align}
  \frac{\partial}{\partial t} ( E_s+E_t)
  &= \kP \rho \left( c a T^4 - \cred (E_s+E_t)\right),\\
  \frac{\partial e}{\partial t} &= \kP \rho 
  \left(\cred (E_s+E_t)
    - c a T^4 \right).
\end{align}

\section{Tests}\label{tests.sec}
We now describe tests of our RHD implementation, focusing on the new
additions. We start with tests of the M1 closure dealing with free
streaming and dust-coupled photons, in \Sec{jiang.sec} and
\Sec{gonzalez.sec}, respectively. Then, in \Sec{dir_press_test.sec},
we analyse the effect of direct radiation pressure from ionising
photons, testing the validity of the momentum transfer from photons to
gas.  In \Sec{diff_test.sec}-\Sec{3d_diff_test.sec} we go on to test
our trapping method for the diffusion of photons in under-resolved
optically thick regimes. Finally, in \Sec{davistest.sec} we test the
full RHD implementation of multi-scattered IR radiation interacting
with dust via momentum and temperature exchange, in an occasionally
optically thick limit, reproducing the recent 2-D experiments of
\cite{Davis:2014jl} on the competition between radiation pressure and
gravity.

\subsection{Free-streaming Radiation from a Thin Disk}\label{jiang.sec}
In \RRT{}, it was demonstrated that while the M1 closure deals well
with single sources of radiation, it fails in-between multiple
sources, creating spurious sources of perpendicular radiation where
opposing radiation flows should more realistically pass through each
other. The point of this first test is to investigate how well the M1
method does in a geometry where we might expect it to fail. We are
inspired here by a similar test which has been performed by Jiang et
al. (in prep.), to compare the behaviour of their Variable Eddington
Tensor closure \citep[VET, e.g.][]{Jiang:2012kp} against M1 and Flux
Limited Diffusion (FLD).

We consider a multiple source geometry which is quite relevant in the
astrophysical context: emission from a thin (galactic) disk,
surrounded by a torus of optically thick gas. We compare, in a 2-D
setup, the converged result of a hydrodynamically static \ramsesrt{}
experiment to an analytically derived result.

The setup is as follows. The simulation box is a square of $1$ cm on a
side, resolved by $128^2$ cells. At $0.1$ cm from the bottom of the
box, centered along the box width, is an emitting horizontal disk, or
line in 2-D, since the disk plane is perpendicular to the simulated
2-D plane. The disk spans one cell in height, and has a length of
$L=0.125$ cm, which corresponds to 16 cell widths. For convenience, we
define the origin to lie at the center of the emitting disk, so the
disk end coordinates are $\pm(L/2,0)$. The disk has a constant energy
density, $E_0$, (imposed in every timestep) of monochromatic radiation
that only interacts with the gas via hydrogen ionisation.

In the background the box contains hot and diffuse ionised gas, while
surrounding the disk is a one-cell high torus, in the same plane as
the disk, of cold and dense neutral gas which is optically thick to
the radiation. The important point is that the background gas is
optically thin, allowing the radiation to pass unhindered, while the
torus instantly absorbs all radiation that enters it, and re-emits
nothing.\footnote{For completeness, the properties of the radiation,
  source, and gas are as follows: the source energy density is
  $E_0=2.2\times 10^{19} \ \edens $, the photon energy is $13.6$ eV,
  and the hydrogen ionisation cross section is $\sigma_{\rm HI}=3
  \times 10^{-18} \ \cs$. The background gas has density $10^{-10} \
  \gcc$ and temperature $10^6$ K, while the torus that surrounds the
  radiation source has density $10^{30} \ \gcc$ and temperature $100$
  K.}

For such a setup, the field morphology can be expressed analytically.
For any point $(x,y)$ in the box, a length element $d\ell$ at location
$(\ell,0)$ along the emitting disk subtends an angle
\begin{align}
d\Omega = \frac{y \ d\ell}{y^2+(x-\ell)^2}.
\end{align}
Assuming isotropic emission and a razor-thin disk, the contribution
from $d\ell$ to the radiation density at $(x,y)$ is
\begin{align}
d E(x,y) = \frac{E_0}{2 \pi} d\Omega.
\end{align}
The energy density at $(x,y)$ can then be obtained by integrating the
contributions from the whole disk:
\begin{align} \label{Jiang.eq} E(x,y) &= \int_{\rm disk} dE(x,y)
  = \int_{-L/2}^{L/2} \frac{E_0}{2 \pi} \frac{y \ d\ell}{y^2+(x-\ell)^2} \\
  &= \frac{E_0}{2 \pi} \left[ \arctan{\frac{L/2-x}{y}} +
    \arctan{\frac{L/2+x}{y}} \right]. \nonumber
\end{align}

\begin{figure}
  \centering
  \includegraphics[width=0.35\textwidth]
    {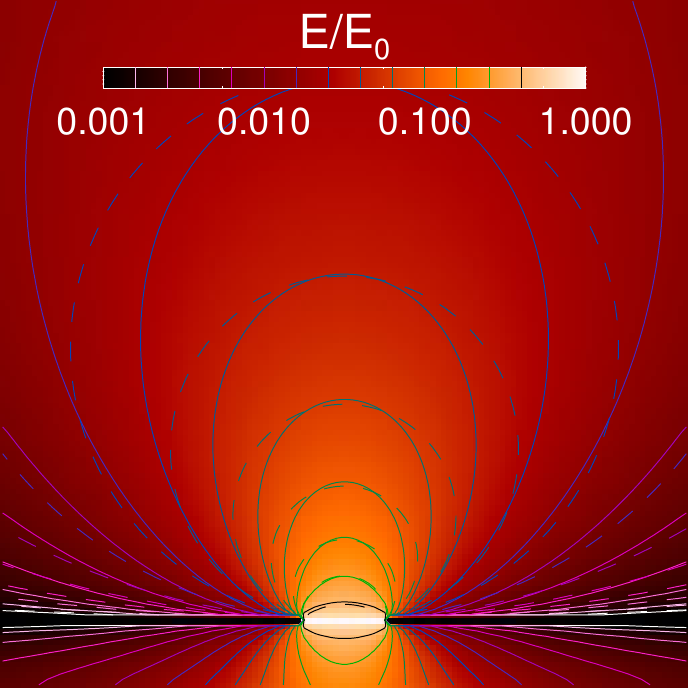}
  \caption
  {\label{Jiang.fig}Emission from a thin disk. The colour scheme and
    solid contours show the radiation density obtained by \ramsesrt{},
    relative to the injected density in the disk at the bottom center,
    while the dashed contours show the exact analytic result from
    \Eeq{Jiang.eq}. The contour values are marked in the colorbar. The
    \ramsesrt{} results agree fairly well with the analytic
    prediction.}
\end{figure}

In \Fig{Jiang.fig} we map the converged radiation density obtained
from \ramsesrt{}, in the color scheme and solid contours, and compare
it to \Eeq{Jiang.eq}, shown as dashed contours. Comparison of the
contours reveals that the M1 scheme does well, though not perfectly,
at reproducing the correct result in this astrophysically relevant
setup. The discrepancy stems from the well-known disadvantage of the
M1 method in dealing with radiation streaming in different directions
in the same point, which results in the radiation being too collimated
perpendicular to the disk. We stress, however, that
\emph{qualitatively}, but not exactly \emph{quantitatively}, the
correct morphology is obtained by \ramsesrt{}.

\subsection{Dust Absorption}\label{gonzalez.sec}
In this test, which is inspired by a similar one from
\cite{Gonzalez:2007gd}, we examine how well the M1 method performs in
producing the correct radiation morphology in the case of absorption
in the optically semi-thick regime. This is again a pure RT test, with
the HD turned off. A 2-D square box $7.48 \times 10^{12}$ cm on a side
is resolved with $64^2$ cells and contains a homogeneous medium with
$\kP \rho=\kR \rho=10^{-12} \, \cmmone$, making the optical depth of
the box $\taub=7.48$. The box is illuminated from the left side by an
incoming horizontal flux of radiation
$F_*=5.44 \times 10^{4} \ \ergs \ \cmmone$. We impose the incoming
radiation by setting a constant $\cred E=F_x=F_*$, and $F_y=0$ in the
left ghost\footnote{\emph{Ghost} cells lie exterior to the box
  boundary on all sides, and define the box boundary conditions. They
  are necessary for the advection in and out of cells interior to the
  box boundaries.} region, and for the remaining three boundaries we
set $E=F_x=F_y=0$. We run until a converged static state has been
reached (which we verified is independent of the light speed used).

The resulting converged gas temperature profile does not depend on the
chosen value for $\kP$, as long as it is nonzero to ensure coupling
between the radiation and gas temperature, and thus eventual
convergence towards $\Tgas=\Trad$ (only the time to reach convergence
depends on $\kP$). The test is thus equivalent to a pure
\emph{scattering} test. We exploit this by comparing the \ramsesrt{}
results to an equivalent setup run with a computation routine,
described in \App{fullRT.app}, that solves the full RT equation
(\ref{RT.eq}) on a four-dimensional grid - with $64^2$ physical
dimensions, and $32^2$ angular bins.  We emit radiation at the rate
$F_*$ in the $x$-direction into the left side of the box, and
otherwise set zero-valued boundaries for the radiation.  The full RT
routine does not evolve (or store) the gas temperature, but is run
instead in pure scattering mode, with the scattering opacity equal to
$\kR$. We compare the \ramsesrt{} gas temperature to the radiation
temperature produced by the full RT routine, which should ideally
converge to the same values.

\begin{figure}
  \centering
  \includegraphics[width=0.35\textwidth]
    {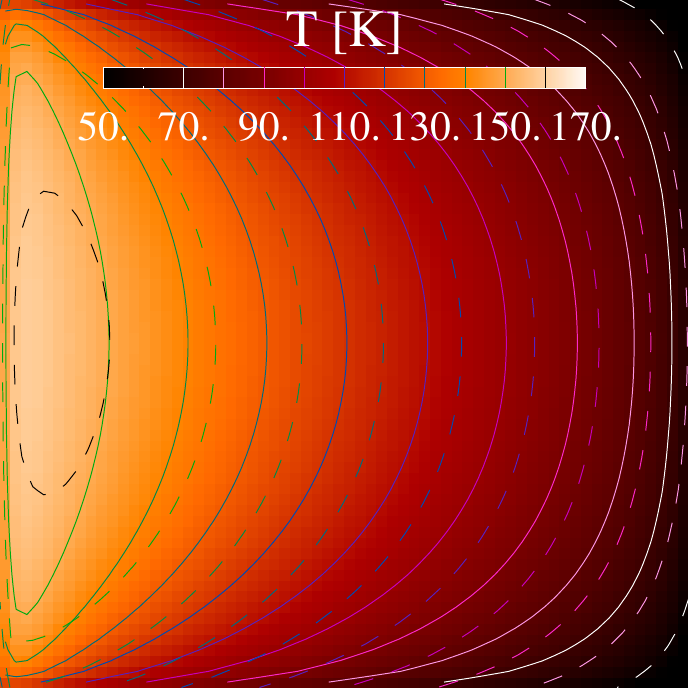}
  \caption
  {\label{Gonz1_test.fig}Two-dimensional photon scattering test, with
    an optical depth from side to side of $\taub=7.48$. The image
    shows the equilibrium state reached in the test. The colors and
    overlaid solid contours indicate the \ramsesrt{} gas
    temperature. For comparison, the dashed contours show results, in
    the form of \emph{radiation} temperature, from an identical test
    run with a full RT code. The results produced by \ramsesrt{} are
    qualitatively similar to the full RT results, but differ in value
    by $10-20\%$. }
\end{figure}

The results are shown in \Fig{Gonz1_test.fig}, where we map with color
and solid contours the gas temperature in \ramsesrt{}. For comparison,
we plot in dashed contours the converged radiation temperature in the
full RT calculation. The results agree well in terms of the shape of
the radiation field, and the accuracy of the \ramsesrt{} produced
radiation field is at the $\sim 10\%$ level compared to the full RT
calculation. The discrepancy can be attributed in part to the M1
moment method directly and its approximative approach to the
collisionless nature of radiation, but in part the boundary conditions
are to blame, which are not exactly equivalent in \ramsesrt{} on one
hand and in the full RT code on the other. The zero-valued boundary
conditions in M1 `suck' radiation out from the top, bottom, and right
sides, while the inwards flux at the right boundary (where the
discrepancy is worst) prevents scattered radiation from flowing back
out of the box.

\subsection{Tests of Direct Pressure from Ionising
  Radiation}\label{dir_press_test.sec}
We aim to demonstrate with the following RHD tests that radiation
pressure in \ramsesrt{} is robustly implemented, i.e. momentum is
correctly deposited from photons to gas. In what follows, we assume an
idealised case of pure hydrogen gas, which is initially homogeneous
and isothermal, and monochromatic photons, and we ignore the effect of
gravity. The setup is a radiation source of luminosity $L$ placed at
the origin in a medium of homogeneous density $\rho_0$ which turns of
at time $t=0$, and we are interested in following the expansion of the
gas due to the direct ionising radiation pressure.  For the tests to
be meaningful, we first need analytic expressions to compare against.

\subsubsection{Analytic expectations} \label{HIIreg_analytics.sec}
\cite{Wise:2012dh} present a simple analytic argument to demonstrate
the effect of radiation pressure in dwarf galaxies. The expression is
derived from requiring momentum conservation in the swept-up gas
around the radiation source, ignoring gravity and thermal pressure,
and describes the radial position $r$ of the expanding density front,
\begin{align}\label{momCons1.eq}
r(t)=\left( \rs^4+2At^2\right)^{1/4}, 
\end{align}
where $A=3L/4\pi\rho_0c$, and $\rs$ is the Str\"omgren radius, at
which an optically thick shell forms at $t\approx 0$\footnote{the
  creation time of the Str\"omgren sphere, which is approximately the
  recombination time, is assumed to be short compared to the
  hydrodynamical response of the gas, an assumption which holds in our
  tests (see \Fig{HIreg_shells.fig}, though it barely holds in the
  highest density case). },
\begin{align} \label{stromgren.eq}
  \rs &= \left( \frac{3L}{4 \pi \recB \nho^2 \ephot} \right)^{1/3} 
  \\
  &= 1.8 \ {\rm pc} \ \left( \frac{L}{\lsun} \right)^{1/3}
  \left( \frac{\nh}{1 \, \cci} \right)^{-2/3}. \nonumber
\end{align}
Here, $\recB$ is the case B recombination rate, which we take to be
equal to $2.5 \times 10^{-13} \ \ccs$, approximately valid in
photo-ionised hydrogen gas, $\nho=\rho_0/m_{\rm p}$ is the hydrogen
number density, $\ephot$ is the monochromatic photon energy, which we
take to be the hydrogen ionisation energy of $13.6$ eV, and we assume
a Solar luminosity of $\lsun = 3.84 \times 10^{33} \ \ergs$ (in
ionising photons).

We will present expanding \hii{} region experiments where we compare
the front position against \Eeq{momCons1.eq}. However, we find \emph{at
  best}, that the simulated expansion only partially follows the
analytic prediction. Firstly, the expansion tends to be dominated by
photoionisation heating, which is not described by
\Eeq{momCons1.eq}. Second, even if the effect of heating is negligible,
the expansion eventually stalls due to thermal gas pressure on the far
side, leaving a semi-stable bubble of diffuse ionised gas surrounded
by a denser neutral gas. The final radius of the bubble is dictated by
the combined effect of photoionisation heating and the direct
radiation pressure.

We can consider separately, for radiation pressure and photo-heating,
roughly how far each of these mechanisms are expected to sweep the
gas.

For the radiation pressure, ignoring the effect of photo-heating, the
bubble will reach a radius $\rrpr$ where the gas pressure outside the
bubble equals the outwards radiation pressure at the surface, i.e.
\begin{align}\label{pressure_competition.eq}
  \nho \kb T_0 = \frac{L}{4\pi \rrpr^2 c},
\end{align}
where $T_0$ is the outer gas temperature and $\kb$ is the Boltzmann
constant. Solving for the bubble radius gives
\begin{align}\label{rad_rtpress.eq}
  \rrpr &= \sqrt{\frac{L}{4 \pi c \nho \kb T_0}} \\
  &=0.28  \ {\rm pc} \ \left( \frac{L}{\lsun} \right)^{1/2}
    \left( \frac{\nh}{1 \, \cci} \right)^{-1/2} 
    \ \left( \frac{T_0}{10^2 \ {\rm{K}}} \right)^{-1/2}. \nonumber
\end{align}

With photoionisation heating dominating, the under-dense bubble is
supported by inner gas pressure, i.e.
\begin{align}\label{Tpress_eq.eq}
  \nin \Tion = \nho T_0,
\end{align} 
where $\nin$ and $\Tion$ are the gas density and temperature inside
the bubble, somewhat incorrectly assumed to be homogeneous, and the
density and temperature outside are just the initial homogeneous
values. Given a radius $\rT$ of the thermally supported bubble, the
ionising luminosity of the central source supports an equal rate of
recombinations in the bubble, i.e.
\begin{align}
  \frac{L}{\ephot} = \frac{4}{3}\pi \rT^3 \recB \nin^2,
\end{align}
From this we can solve for the gas density inside the bubble, which we
insert into \Eeq{Tpress_eq.eq}, giving
\begin{align} \label{rad_therm.eq}
  \rT &=  \left( \frac{\Tion}{T_0} \right)^{2/3} \rs \\
  &= 39 \ {\rm pc} \ \left( \frac{L}{\lsun} \right)^{1/3}
  \left( \frac{\nh}{1 \, \cci} \right)^{-2/3} \nonumber \\
  & \ \ \ \ \ \ \ \ \ \ \ \
  \left(\frac{\Tion}{10^4 \, {\rm K}} \right)^{2/3}
  \left(\frac{T_0}{10^2 \, {\rm K}} \right)^{-2/3}. \nonumber
\end{align}

We can now compare the radius of the radiation pressure supported
bubble versus the radius of the thermally supported bubble. The
condition for radiation pressure to start dominating over
photoionisation heating is 
\begin{align}
  \rrpr > \rT.
\end{align}
Substituting equations \eq{stromgren.eq}, \eq{rad_rtpress.eq}, and
\eq{rad_therm.eq} then gives the condition
\begin{align} \label{rtpress_cond.eq}
  L  & > \frac{1}{\nho}\frac{\Tion^4}{T_0} \frac{36 \pi c^3 
    \kb^3}{\recB^2 \ephot^2} \\
  & = 7 \times 10^{12} \,  \lsun \ 
  \left(  \frac{\nho}{1 \, \cci}  \right)^{-1}
  \left(  \frac{T_0}{10^2 \, \rm{K}} \right)^{-1} \nonumber \\
  & \ \ \ \ \ \ \ \ \ \ \ \ \ \ \ \ \ \ \ \ 
    \left( \frac{\Tion}{10^4 \rm{K}} \right)^4
  \left( \frac{\ephot}{13.6 \, \rm{eV}} \right)^{-2}.
  \nonumber
\end{align}
Admittedly, a range of assumptions and approximations go in, but
\Eeq{rtpress_cond.eq} nevertheless gives an idea of the luminosities
required for ionising radiation pressure to give a strong boost over
the effect of photoionisation heating. Clearly \textit{both} large
luminosities and gas densities are required for this to
happen. However, the relative difference in the equilibrium radii
scales only very weakly with the density and luminosity, i.e.
\begin{align}
  \frac{\rrpr}{\rT} \propto (L \, \nho)^{1/6},
\end{align}
so even if the condition of \Eeq{rtpress_cond.eq} is far from met,
radiation pressure may well give a modest boost to the thermally
driven expansion. Conversely, this also means that a prodigious
luminosity and/or density is required for the photoionisation heating
to become negligible, as is generally acknowledged in the literature
\citep[see][and references therein]{Krumholz:2009iu}.

We can also consider the relevant physical scales for ionising
radiation pressure by requiring that it is stronger than the thermal
pressure in a \stromgren{} sphere,
\begin{align}
  \frac{L}{4 \pi c \rS^2} > \nho \kb \Tion.
\end{align}
Solving directly for the luminosity gives \Eeq{rtpress_cond.eq} with
the outer temperature, $T_0$, removed. But for the physical scale, we
can instead use \Eeq{stromgren.eq} to eliminate $\nho$, giving the
requirement on the \stromgren{} radius that
\begin{align}\label{rtpress_cond2.eq}
  \rS & < \frac{\recB}{12 \pi c^2 \kb^2} \frac{L \ephot}{\Tion^2}  \\
  & = 0.1 \, {\rm pc } \, 
  \left( \frac{L}{10^6 \, \lsun} \right)
  \left( \frac{\ephot}{13.6 \, {\rm eV}} \right)
  \left( \frac{10^4 {\rm K}}{\Tion} \right)^2.
  \nonumber
\end{align}
Comparing with \Eeq{stromgren.eq}, this translates to a young stellar
population of $\approx 10^3 \ \Msun$ ($L\approx 10^6 \, \lsun$),
embedded in gas with $\nho \sim 10^5 \ \cci$, which is currently
beyond, but not far from, the resolution limits of most galaxy-scale
simulations.

\subsubsection{Expanding HII regions}
\begin{table}
  \centering
  \caption
  {Expanding \hii{} region tests. All tests are run in a square box
    with $128^3$ cells, with a source luminosity of $10^6 \ \lsun$,
    a monochromatic photon energy of $15$ eV, and a reduced speed
    of light factor $\fc=10^{-3}$. The columns list, from left to
    right, the initial homogeneous gas number density, $\nho$, the expected
    thermally supported bubble radius, $\rT$, direct radiation pressure
    supported bubble radius, $\rrpr$, the box width, $\Lbox$, the
    run time of each test, $\trun$, and, for comparison, the
    recombination time $\trec=(\nho \recB)^{-1}$, which is
    approximately the time it takes for the Str\"omgren sphere to develop.}
  \label{wise.tbl}
  \begin{tabular}{m{0.5cm} ccccc}
    \toprule
    $\nho$ & $\rT$ & $\rrpr$ & $\Lbox$ & $\trun$ & $\trec$    \\ 
    $[\cci]$ & [pc] & [pc] & [pc] & [Myr] & [Myr]    \\ 
    \midrule
    $10^0$  & $291$ & $36$ &  $450$ & $10^3$ & $10^{-1}$ \\
    $10^3$  & $2.9$ & $1.1$ &  $5.5$  & $10$ & $10^{-5}$ \\
    $10^5$  & $0.13$ & $0.11$ &  $0.3$  & $0.3$ & $10^{-6}$\\
    $10^7$  & $6 \times 10^{-3}$ & $11 \times 10^{-3}$ 
            & $2 \times 10^{-2}$ & $10^{-2}$   & $10^{-8}$ \\
    $10^9$  & $2 \times 10^{-4}$ & $11 \times 10^{-4}$ 
            & $1.4 \times 10^{-3}$ & $10^{-3}$  & $10^{-10}$ \\
    \bottomrule
  \end{tabular}
\end{table}

\begin{figure*}
  \centering
  \includegraphics[width=\textwidth]
  {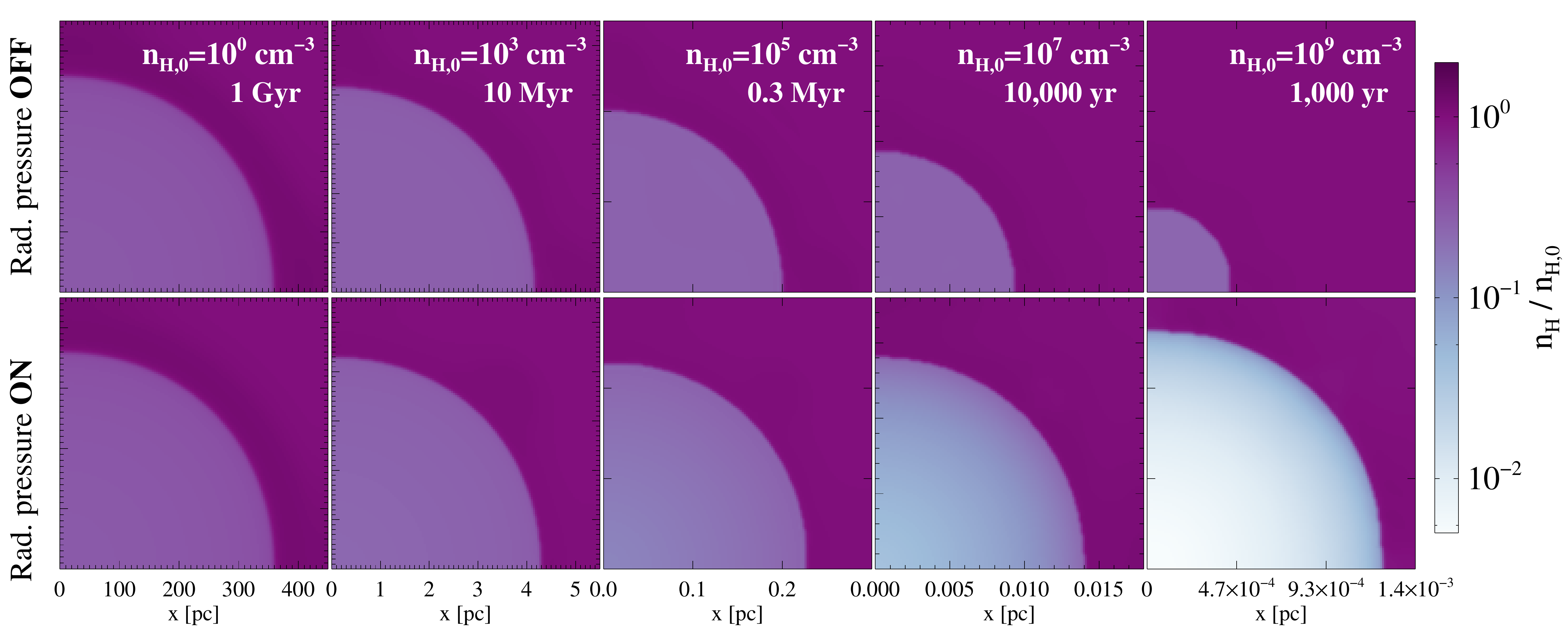}
  \caption{\label{HIIreg_maps.fig}Slices of the simulation box, on the
    side containing the radiation source, showing gas density,
    normalised to the initial density, at the end of the expanding
    \hii{} region tests. The upper row shows tests with direct
    ionising radiation pressure turned off, i.e. photoionisation
    heating only, and the lower row shows the corresponding runs with
    the radiation pressure turned on. The panels are ordered from left
    to right by the initial homogeneous gas density, as indicated in
    the top right corner of the upper row panels.}
\end{figure*}

We set up a square 3-D box and place in the corner a source of
luminosity $L=10^{6} \ \lsun$, emitting monochromatic ionising photons
with energy $\ephot=15$ eV\footnote{Average ionising photon energies
  from young stellar populations are larger by a few eV. However, we
  use a low photon energy to minimise photoionisation heating and give
  radiation pressure a head start, as higher photon energies increase
  the heating rate in the \hii{} region. } ($1.8\times 10^{50}$
photons per second) and hydrogen ionisation cross section
$\csH=3\times 10^{-18} \ \cs$, into an initially homogeneous neutral
pure hydrogen gas (no helium, metals or dust) at a temperature of
$10^4$ K. The box boundaries adjacent to the source are reflective and
the opposite sides have outflow boundaries. We use $128^3$ cells, and
reduce the speed of light by a factor $\fc=10^{-3}$. Even at this low
light speed the run-time is hundreds of light-crossing times in each
run, so this has no effect on the later stages of development.

\begin{figure*}
  \centering
  \subfloat[$\nho=10^{3} \, \cci$]{\includegraphics[width=0.4\textwidth]
    {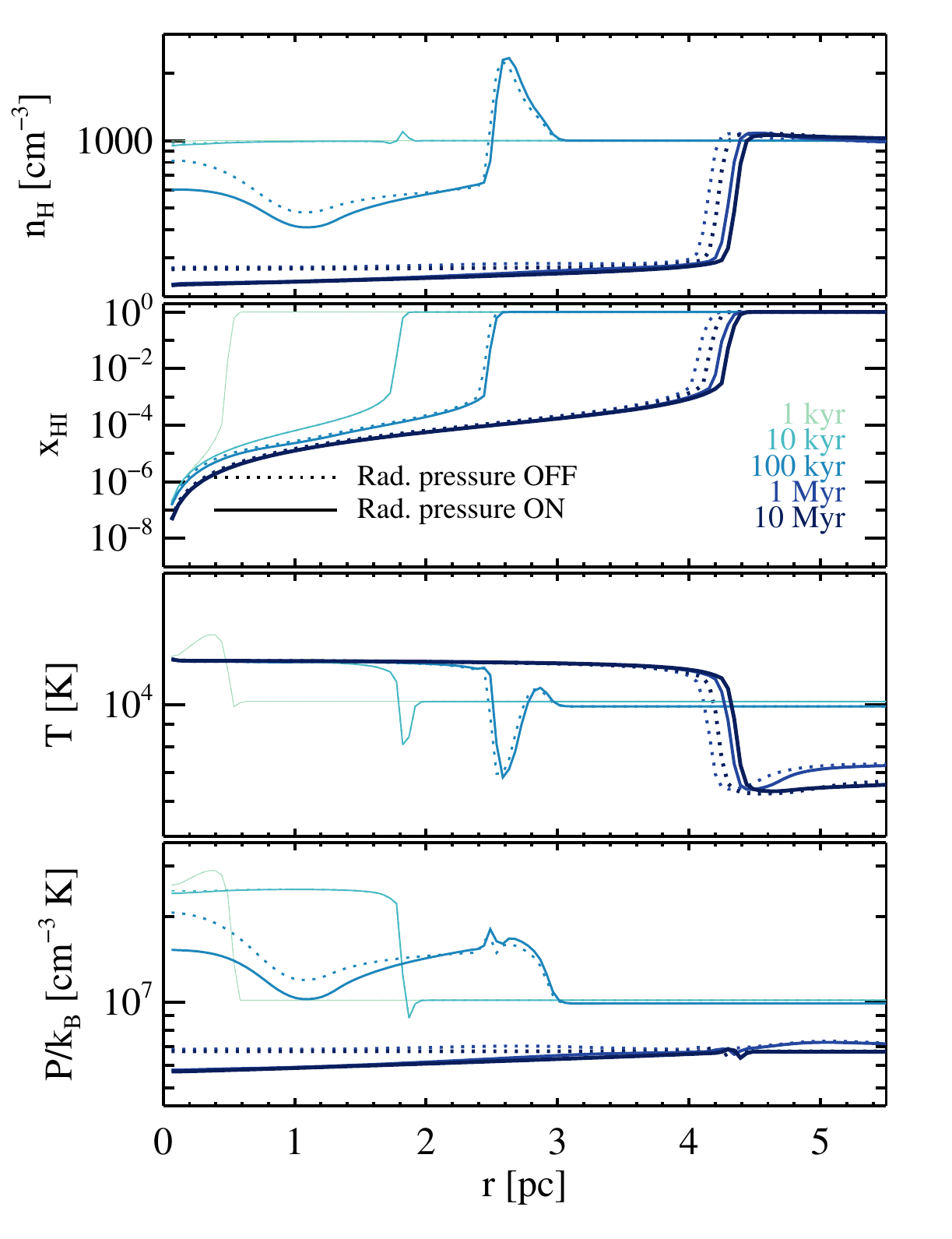}} 
  \subfloat[$\nho=10^{9} \, \cci$]{\includegraphics[width=0.4\textwidth]
    {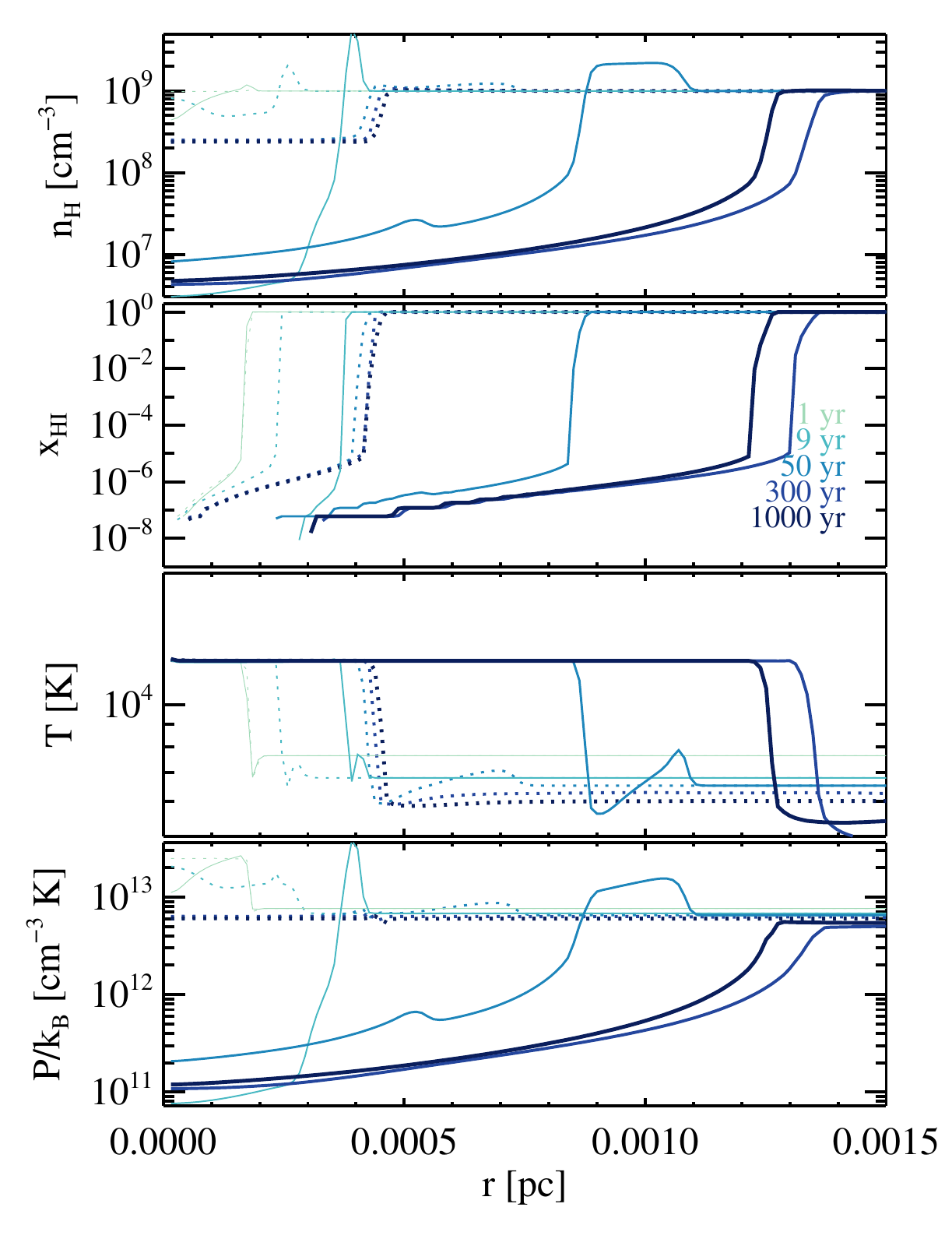}}
  \caption
  {\label{HIIreg_profiles.fig}Radial profiles of, from top to bottom,
    the gas density, neutral fraction, temperature, and gas pressure,
    for the expanding \hii{} region tests, with a $10^6 \, \lsun$
    source radiating ionising photons into an initially homogeneous
    neutral medium. The plots to the left show the case with
    $\nho=10^{3} \, \cci$, where radiation pressure has only a
    marginal effect compared with photoionisation heating, and the
    plots to the right show $\nho=10^{9} \, \cci$, where radiation
    pressure dominates over photoionisation heating. Runs with only
    photoionisation heating are represented by dotted curves, while
    runs that in addition include direct pressure from the ionising
    photons are represented by solid curves. The curve colors (and
    thickness) represent the profile times, as indicated in the
    ionisation fraction plots.}
\end{figure*}

To compare regimes where either ionisation heating or ionisation
pressure dominates, we compare sets of runs at five different
initial densities $\nho$, presented in \Tab{wise.tbl}. For each
initial density we run two tests: with and without direct radiation
pressure.  The table also shows the run time ($\trun$), the box width
($\Lbox$), and our estimates for the thermally supported bubble radius
($\rT$, \Eq{rad_therm.eq}) and the direct radiation pressure supported
radius ($\rrpr$, \Eq{rad_rtpress.eq}), where we have used a bubble
temperature of $\Tion=1.3 \times 10^4$ K and an external temperature
of $T_0=6 \times 10^3$ K, based approximately on the temperature
profiles in the end results (see \Fig{HIIreg_profiles.fig}: the
radiation heats the ionised gas, and the shielded neutral gas
eventually cools due to residual collisional ionisation). Comparing
the $\rT$ and $\rrpr$ values in the table, photoionisation heating
should dominate in the test with the lowest initial density,
$\nho=1 \, \cci$, but with higher densities radiation pressure should
have an increasing effect, and should dominate at the highest initial
density of $\nho=10^9 \, \cci$.

\Fig{HIIreg_maps.fig} shows slices, at the side of the box containing
the radiation source, of gas density at the end of each run. Comparing
the maps with and without direct radiation pressure, i.e. the upper
versus lower row of maps, it is clear that radiation pressure has a
negligible effect at the lowest initial densities, while it gradually
overtakes the effect of photo-ionisation heating at higher gas
densities. It can also be seen that radiation pressure, once it
becomes effective, is more efficient at driving the gas out of the
bubble, creating much lower internal densities than with
photoionisation heating only.

\begin{figure}
  \centering
  \includegraphics[width=0.48\textwidth]
  {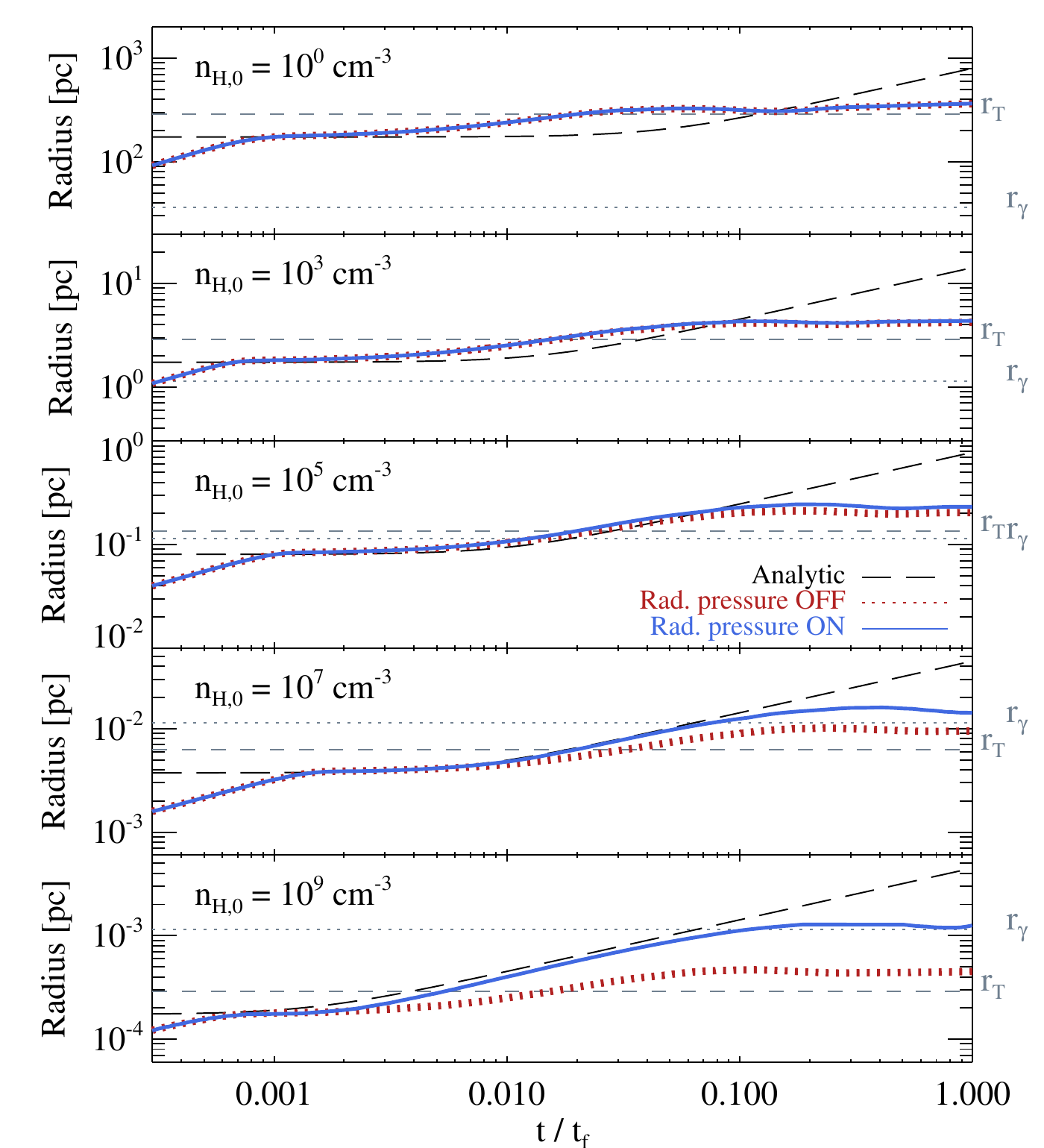}
  \caption{\label{HIreg_shells.fig}Evolution of radiation-powered
    \hii{} region radius (the radius at which the ionised fraction is
    $0.5$), for increasing initial gas density (top to bottom
    plot). In each plot, the solid blue (dotted red) curve shows the
    bubble radius with direct radiation pressure turned on (off), and
    the black dashed curve shows the analytic expectation from
    momentum conservation (\Eq{momCons1.eq}). Dashed grey horizontal
    lines show the expected thermally supported bubble radius ($\rT$,
    \Eq{rad_therm.eq}), while dotted grey horizontal lines show the
    expected radiation pressure supported radius ($\rrpr$,
    \Eq{rad_rtpress.eq}), where we have used a bubble temperature of
    $\Tion=1.3 \times 10^4$ K and an external temperature of
    $T_0=6 \times 10^3$ K, based approximately on the temperature
    profiles in the end results (see \Fig{HIIreg_profiles.fig}). As
    those simple analytic estimates predict, photoionisation heating
    dominates at the lower densities, but radiation pressure starts to
    take over at high densities, with an expansion towards the final
    bubble radius that is well described by momentum conservation. The
    early deviations from the analytic results, at
    $t\la 0.001 \, \trun$, correspond to the ionisation front
    expansion towards the \stromgren{} radius, which in the analytic
    arguments was assumed to happen instantaneously.}
\end{figure}

\Fig{HIIreg_profiles.fig} shows radial profiles of, from top to
bottom, gas density, neutral fraction, temperature and thermal
pressure, taking average values in radial bins from the source. We
show profiles for two sets of initial densities, one at which
radiation pressure is just starting to have an effect
($\nho=10^3 \, \cci$, left panel), and the highest initial density, at
which radiation pressure clearly dominates ($\nho=10^9 \, \cci$, right
panel). The density profile plots (top) show how shells of over-dense
gas are ejected from the ionisation-front, leaving behind a
semi-stable bubble of diffuse gas.  For the lower-density case (left
panels), the profiles with/without radiation pressure are quite
similar. The addition of radiation pressure only slightly advances the
bubble and yields a slightly lower density and gas pressure at the
bubble center. We note that a similar comparison of profiles at the
lowest initial density, $\nho=1 \, \cci$, reveals negligible
differences between the runs with radiation pressure on or off (not
shown), so we are indeed considering densities where radiation
pressure is just beginning to have a non-negligible effect compared
to photoionisation heating.

For the high density case (right panels in \Fig{HIIreg_profiles.fig}),
turning on the radiation pressure has a very substantial
effect. Compared to the photoionisation heating only case, both the
inner bubble density and pressure are almost two orders of magnitude
lower, while the temperature remains nearly unchanged. The bubble is
now mostly supported by direct radiation pressure, as can be clearly
seen by comparing the thermal pressure profiles (bottom left
plot). With only photoionisation heating the bubble is supported by
thermal pressure, which is identical inside and outside the
bubble. With radiation pressure turned on, the thermal pressure drops
dramatically inside the bubble and the direct radiation pressure
compensates to maintain the large steady bubble, such that the
\emph{sum} of gas and radiation pressure is identical on each side of
the interface. 

Finally, \Fig{HIreg_shells.fig} shows the expansion of the ionisation
front (I-front, which we define to be at $\xhi=0.5$), which here is a
proxy for the radius of the under-dense bubble, in each of the runs,
with the plots ordered by increasing density from top to bottom. We
show the I-front expansion as predicted by analytic momentum
conservation (\Eq{momCons1.eq}, dashed black), and from the runs, with
photoionisation heating only (dotted red) and with added direct
radiation pressure (solid blue). Grey lines show our estimate of the
radiation pressure supported radius $\rrpr$ (\Eq{rad_rtpress.eq},
dotted), and the thermally supported radius $\rT$ (\Eq{rad_therm.eq},
dashed), given in \Tab{wise.tbl}. If the numerical I-front expansion
is regarded closely, it can be seen that the front overshoots slightly
in all runs, due to the momentum of the expanding gas, and then
backtracks to reach a radius where the inner and outer pressure is in
equilibrium. This effect can also be seen in the right panel of
\Fig{HIIreg_profiles.fig}, if the curves for $3 \times 10^2$ and
$10^3$ years are compared.

Two important points can be inferred from
\Fig{HIreg_shells.fig}. Firstly, the numerical experiments roughly
reproduce the analytic expectations, laid out in
\Sec{HIIreg_analytics.sec}, for the relative roles of photoionisation
heating and direct radiation pressure. For the lowest initial density
(top plot), the bubble radius $\approx \rT$, while at the highest
density (bottom plot) it goes out to $\approx \rrpr$. The second point
is that when radiation pressure dominates the bubble expansion,
\emph{and while the bubble is expanding towards its final radius}, the
momentum conserving prediction, \Eeq{momCons1.eq}, is reproduced by the
numerical results (bottom plot)\footnote{The analytic result is not
  reproduced at the very start, at $t\la 0.001 \, \trun$. This is the
  I-front expansion towards the \stromgren{} radius, ignored in the
  arguments leading to \Eeq{momCons1.eq}, and during which the gas
  density stays more or less constant.}.

All in all, these results strongly indicate that \ramsesrt{} correctly
models direct radiation pressure \emph{and} photoionisation
heating. As a further validation, the results are qualitatively in
good agreement with the numerical experiments of \cite{Sales:2014gw},
where ionising radiation pressure begins to dominate over
photoionisation heating at similar luminosities and densities as in
our case (see their Figure 6).

\subsection{Resolved versus unresolved photon
  diffusion}\label{diff_test.sec}
We will show quantitative tests of photon trapping in the next
subsections, but we shall start with a simple demonstration of how it
produces robust results when the mean free path is unresolved.

We consider a simple 2-D pure RT test, i.e. with the HD turned
off. The box contains a homogeneous medium which is optically thick to
IR radiation, with an optical depth of $\taub=200$. Through the left
boundary we emit a constant IR flux of
$5.44 \times 10^4 \ \ergs \ \cmmone$. The remaining sides of the box
have zero-value boundaries. We use a full light speed, but note that
the results are independent of the light speed used.

\begin{figure}
  \centering
  \includegraphics 
    {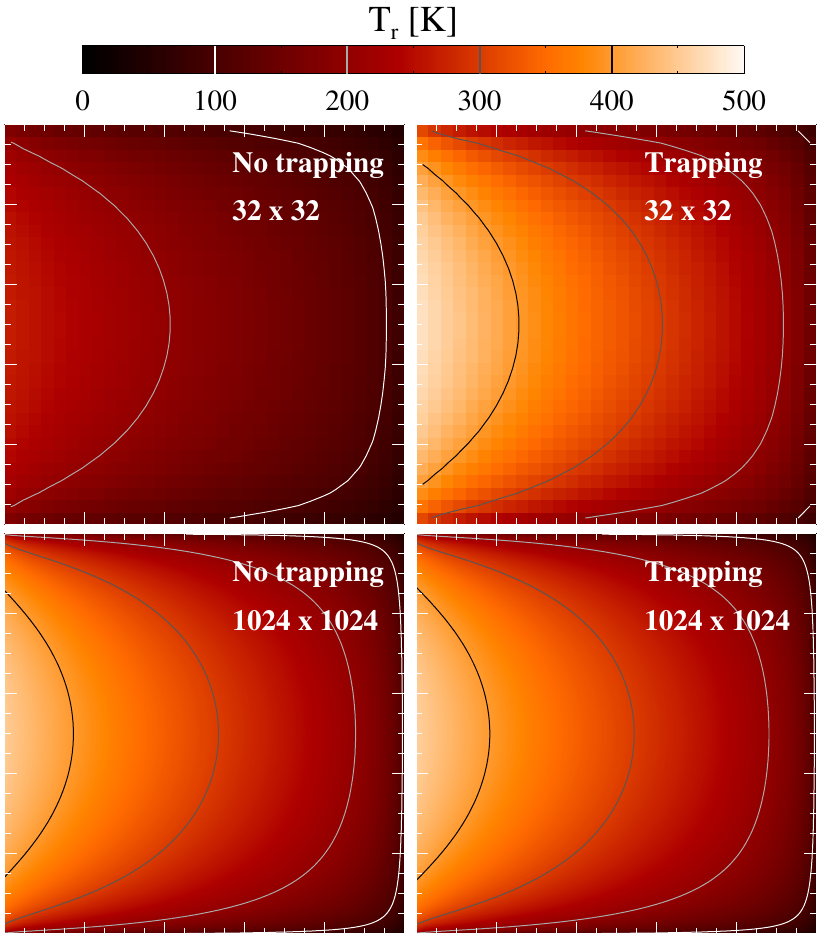}
  \caption
  {\label{diff_test.fig}A demonstration that our method for photon
    trapping produces robust results in an optically thick medium,
    with $\taub=200$. All maps show time-converged results from 2-D
    \ramsesrt{} runs, with a constant flux of photons into the box
    from the left. The color represents the radiation temperature,
    $\Trad$, as indicated by the color bar, and contours mark
    centennial values, also marked in the color bar. The {\bf top
      left} map shows the results without photon trapping in a
    low-resolution run, $32^2$ cells. The {\bf bottom left} map shows
    an identical run, i.e. no trapping, with a much higher resolution
    of $1024^2$ cells. The results are different, since the mean free
    path is resolved by $\approx 5$ cell widths in the high-resolution
    run, whereas a cell width contains $\approx 6$ mean free paths in
    the low-resolution run.  In the {\bf right column} of maps we show
    the results of running with the same pair of resolutions, but with
    photon trapping activated. With photon trapping on, the results
    are much better converged with resolution.}
\end{figure}

We use this setup in four \ramsesrt{} experiments, each running until
a steady-state is reached. We run with a low resolution of $32^2$
cells and a high resolution of $1024^2$ cells, such that the mean free
path is $0.16$ and $5.12$ cell widths, respectively. For each
resolution, we run with and without photon trapping activated.

Without trapping, we should expect more or less correct results in the
high-resolution run, where the mean free path is well resolved, but
incorrect results in the low-resolution run, where the photons diffuse
artificially between the optically thick cells.  With trapping turned
on, photon diffusion is also handled on unresolved scales, and there
should ideally be no difference between the high- and low-resolution
runs (on scales larger than the low-resolution cell width). The
low-resolution results with trapping should resemble those of the
high-resolution run without (and with) trapping.

This is indeed the case, as shown in \Fig{diff_test.fig}, where we map
the steady-state radiation temperature, $\Trad=(E/a)^{1/4}$, in the
four runs. Comparing the low- and high-resolution runs without
trapping (top left and bottom left, respectively), we see a large
qualitative difference in the steady-state radiation field. With the
unresolved mean free path, the photons diffuse numerically from the
optically thick cells, and there is much less buildup of radiation
compared to the higher resolution case, where numerical diffusion is
negligible. Comparing instead the two runs with trapping turned on
(top and bottom right), we find similar results, even if the cell
widths differ by more than an order of magnitude. Furthermore, the
results with photon trapping are also similar to the high-resolution
case without trapping, indicating strongly that the photon trapping
method i) reproduces the correct results when the mean free path is
unresolved, and ii) converges to the correct result when the mean free
path becomes well resolved.

The agreement is not perfect, as can be seen from a careful comparison
of the contours and the box edges. This disagreement stems partly from
the fact that the non-trapping result is still not quite resolution
converged, but more importantly, with trapping turned on, the box
boundary does not behave in the same way along optically thin cells as
it does along optically thick ones. In the optically thin limit (lower
right), the photons freely escape along the boundaries on scales
shorter than the mean free path, and accurately so, since the
boundaries are zero-valued. However, when the mean free path is not
resolved (upper right), the escape of photons along the boundary is
suppressed by the trapping, which essentially assumes the same mean
free path everywhere within the cell, resulting in larger values for
the radiation temperature.

\subsection{Diffusion of a Radiation Flash in 2-D}
\label{2d_diff_test.sec}
We now test whether our implementation of radiation trapping agrees
with analytic expectations of diffusing radiation. We consider two test
cases, in this and the next subsection. In both cases, HD is turned
off.

The first test is a 2-D version of the 1-D test described in
\cite{Commercon:2011eq}. The simulation box is a $1$ cm wide square
composed of $128^2$ gas cells, which contain a homogeneous medium with
$\kR \rho=10^3 \ \cmmone$ (i.e. $\taub=10^3$).  The box is initially
empty of radiation, except for ${\mathcal N}_{0}=10^5$ photons that are
distributed uniformly over four cells at the center of the box, at
which we define the origin of our coordinate system. We then turn on
the RT, allowing the photons to diffuse out of the box. For the
boundary conditions, we apply linear extrapolation to all the RT
variables, from a buffer of two cells inside the border, to determine
the values in ghost cells outside the border. We run this test with
the full light speed, i.e. with $\cred=c$.

\begin{figure}
  \centering
  \includegraphics 
  {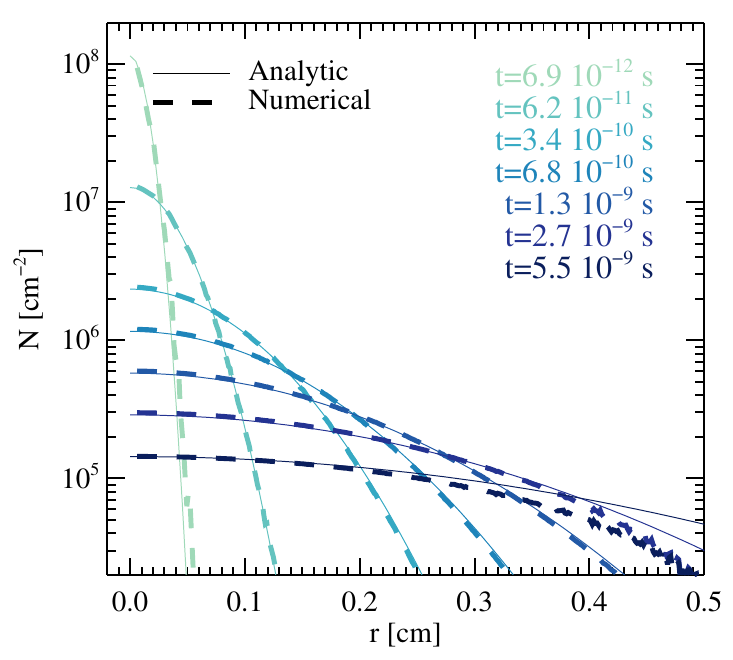}
  \caption{\label{Comm_test.fig}Two-dimensional flash diffusion
    test. Each set of solid (analytic solution) and dashed (numerical
    solution) curves represents the radial radiation profile at the
    time given by the line color, as indicated in the legend. Except
    near the boundary of the box ($r=0.5$ cm), the numerical and
    analytic results agree well.}
\end{figure}

The evolution, with time $t$ and radius $r$ from the origin, of the
photon number density $N$, is given by \citep{Commercon:2011eq}
\begin{align}\label{Comm_ana.eq}
 N(r,t) = \frac{{\mathcal N}_{0}}{2^p \left( \pi \chi t
   \right)^{p/2}} \ {\rm e}^{-\frac{r^2}{4 \chi t}},
\end{align}
where $\chi=c/(3 \kR \rho)$, and $p=2$ is the number of
dimensions. \Fig{Comm_test.fig} shows the time-evolution of the
analytic radiation density profile (solid curves), and compares it to
the test results (dashed), up to $5.5 \times 10^{-9}$ seconds, which
corresponds to $165$ box crossing times in the free-streaming
limit. The numerical results show the sum of the trapped \emph{and}
free-streaming photons (see \Eq{trappedTot.eq}). The agreement is
excellent. The main discrepancy, at the box edges at late times is
caused by the boundary conditions, which release the photons too
efficiently.

We note that we also ran the test with a reduced light speed
$\cred=c\times \fc=c/100$, reproducing exactly the former results, if
the replacement $c\rightarrow\cred$ is made in \Eeq{Comm_ana.eq}, and
the profiles are plotted at the times $t / \fc$, where $t$ is the
profile times in \Fig{Comm_test.fig}. In other words, reducing the
speed of light simply slows the diffusion speed by a factor $\fc$.

We also ran the test with ten times higher and lower optical depth
(via $\kR$). At the higher optical depth, the numerical results come
even closer to the analytic ones. Conversely, at the lower optical
depth, the results visibly diverge from \Eeq{Comm_ana.eq}, as should
be expected in the free-streaming radiation limit.

\subsection{Diffusion of Constant Luminosity Radiation in
  3-D} \label{3d_diff_test.sec}
\begin{figure*}
  \centering
  \includegraphics[width=\textwidth]
  {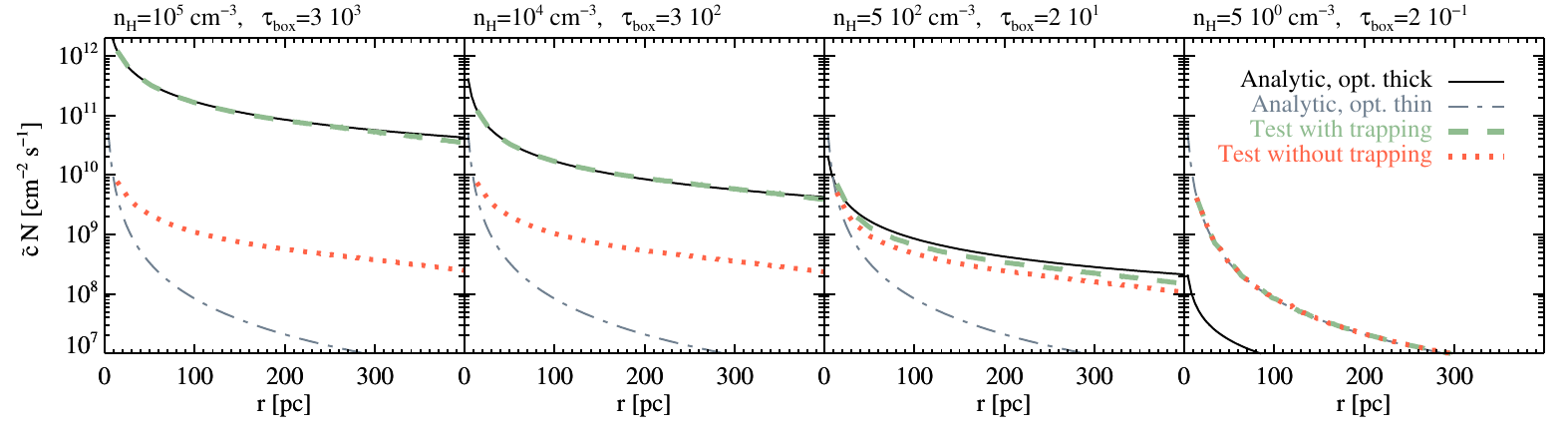}
  \caption{\label{fld.fig}Test of radiation diffusion in a medium of
    decreasing optical thickness (from left to right). The plots show
    time-converged radiation profiles from the source at the center of
    the box in radiation tests with and without trapping (green dashed
    and red dotted lines, respectively). The solid black lines show
    the analytic solution to the diffusion equation,
    \Eeq{fld_test2.eq}, which the tests with trapping should reproduce
    in optically thick gas. The grey dot-dashed lines show the
    analytic solution for free-streaming radiation, which the tests
    should reproduce for vanishing optical thickness, regardless of
    whether trapping is turned on or off.}
    \label{trapping_test}
\end{figure*}

We now consider again radiation diffusion with the hydrodynamics
turned off, but in 3-D, and with a constant luminosity source. We use
a setup, which is relevant for cosmological simulations in terms of
the source luminosity, gas density, metallicity, and spatial
resolution. We put a source with a luminosity $L=10^{50} \ \emrate$
into the center of a box which is resolved by $32^3$ cells, and allow
the radiation to propagate through the homogeneous gas with the
trapping model presented in \Sec{trapping2.sec}, assuming an opacity
$\kR = 10 \ \cmsqg$, until a converged steady-state has been
reached. The box width is $\Lbox = 500$ pc, which gives a cell size of
$15.6$ pc. We then run variants of this setup with varying gas
density, spanning $\nh=5-10^5 \ \cci$, corresponding to optical depths
(through the box) of $\taub \approx 0.2-3 \times 10^3$.

We compare the converged, steady-state, numerical radiation density
profile, as a function of distance from the source, to an analytic
expression which is derived as follows.

In a homogeneous optically thick medium of density $\rho$ and
emittance ${\mathcal L}$ (i.e. luminosity per volume), the local
photon number density, $N$, is described by the diffusion equation,
\begin{align}\label{fld_test0.eq} 
  \frac{\partial N}{\partial t} 
  - \frac{\cred}{3 \kR \rho} \nabla^2 N + {\mathcal L} = 0.
\end{align}
In the steady-state limit, this reduces to the Poisson equation,
\begin{align}\label{fld_test1.eq} 
  \frac{\cred}{3 \kR \rho} \nabla^2 N ={\mathcal L}.
\end{align}
In three dimensions, assuming a single point source of radiation, the
solution is
\begin{align}\label{fld_test2.eq} 
  N(r)=\frac{3\rho\kR L}{4 \pi \cred r},
\end{align}
where $r$ is the distance to the radiation source, and $L$ is the
point source luminosity. \Eeq{fld_test2.eq} is the analytic expression
we can compare to our numerical results.

The analytic argument leading to \Eeq{fld_test2.eq} essentially
assumes infinity in both space and time, i.e. there are no boundaries
or `box' limits, and steady-state can thus only be reached in an
infinite time. For time, we simply run the tests until they converge
to a final solution, but to approximate the infinite spatial
dimensions, we set up the boundaries of the box to roughly match the
expected slope given by \Eeq{fld_test2.eq}\footnote{In the tests with
  the most optically thick gas, free-flow boundary conditions result
  in an overestimate of the radiation in the box, since the gradient
  at the box edge is zero, giving too much back-flow of radiation from
  the boundaries, while zero-valued boundaries give an underestimate
  because the gradient is infinite, and hence no back-flow comes from
  the boundaries.}. The boundary condition for this test is thus
\begin{align}\label{bc.eq} 
  \state_0 = \state_1 
  \left( 1 - \frac{\dx}{\Lbox} \right),
\end{align}
where $\state=\left( \bF, N \right)$ is a cell state, $\dx$ is the
cell width at the boundary, and the subscripts 0 and 1 refer to the
ghost cell and the boundary cell inside the computational domain,
respectively. The boundary can only approximately `mimic' the infinite
space assumption, since the box has a square shape.

\Fig{fld.fig} shows the results of the diffusion tests, where we have
run with a reduced speed of light, $\cred=c/200$. The steady-state
limit for radiation flux is the same as with a full light speed, but
it takes longer, by s factor $\fc^{-1}$, to reach that state. From
left to right, the plots in \Fig{fld.fig} show the test results for
the different gas densities, which translate to different optical
depths. In each plot, the grey dash-dotted lines show the
$N\propto r^{-2}$ profile expected for free-streaming radiation, while
the solid black lines show the optically thick prediction made by
\Eeq{fld_test2.eq}. The dashed green curves show the converged test
results where photon trapping is applied. For comparison, the dotted
red curves show the converged results of identical tests where photon
trapping is deactivated.

In the optically thick case (leftmost two plots), the radiation
profile evolves towards the correct diffusion solution when trapping
is included. On close inspection it can be seen that the test results
(green dashed) do not perfectly follow the analytic prediction near
the edge of the box, but this is purely due to the boundary
conditions, which as we remarked are not correct everywhere due to the
geometry of the box. If the slope at the boundaries is steepened, the
agreement with the analytic result becomes better at $r \approx 250$
pc, where the edge of the box is closest, but at the same time it
becomes worse at $r \approx 350$ pc, corresponding to the box corners,
where the gradient should be shallower.

The third plot from the left shows worse agreement with the analytic
solution, but here the gas is also coming close to the optically thin
regime, and \Eeq{fld_test2.eq} no longer holds. In the rightmost plot
we have the situation where $\taub \ll 1$, and the results agree
with the free-streaming limit, regardless of whether trapping is
turned on or off.

The curve without trapping assumes the correct $\propto r^{-1}$ shape
where $\taub>1$ due to the scattering which isotropises the
radiation in every cell, but the curve fails to follow the correct
scaling with increasing $\taub$.

Again we find that our scheme for trapped radiation
(\Sec{trapping2.sec}) robustly reproduces analytic expectations. We
ran this test as well with an alternative version of our method for
handling the optically thick regime, suggested in \Sec{godflux.sec},
where instead of splitting the photons into trapped and
free-streaming, we apply directly a diffusion operator
$\alpha(\tauc)=(1+3/2 \ \tauc)^{-1}$, where $\tauc$ is the cell
optical depth, to the GLF intercell flux function, as in
\Eq{GLF_mod1.eq}. The results using this alternative version were
identical to using the trapped/streaming photons scheme, which is no
surprise, since the trapped/streaming split essentially amounts to the
same thing for the intercell flux. However, the trapped/streaming
scheme has the further advantages of the trapped photons moving with
the gas, and of a natural inclusion of radiation pressure in the
optically thick regime, neither of which is an issue in this test.

\subsection{Levitation of Optically Thick Gas}\label{davistest.sec}
As a final test of radiation pressure, the radiation-temperature
coupling, multi-scattering, and photon trapping, we repeat the 2-D
experiment described by \cite{Krumholz:2013fa} and
\cite{Davis:2014jl}, hereafter \KT{} and \SD{}, respectively, which
explores the competition between gravity and radiation pressure.

The experiment is interesting in the context of radiation feedback,
because it gives insight into how gravitationally bound gas responds
to multi-scattering radiation pressure. The setup, which represents a
stellar nursery or the central plane of an optically thick galactic
disk, consists of a thin bottom layer of gas, kept in place by
gravity, which is then exposed to an opposing flux of IR radiation.
Even though the radiation flux is sub-Eddington, the effect of
multi-scattering may still lift the gas if the radiation is efficiently
trapped by the gas. However, radiative Rayleigh-Taylor instabilities,
if they develop, suppress the radiation pressure by creating
`chimneys' through which the radiation may escape without efficiently
coupling to the gas.

\KT{} ran the experiment using the flux-limited diffusion (FLD)
method, which essentially solves \Eeq{Flux_asymp.eq}, while making
sure the radiation does not surpass the speed of light in the
optically thin limit. They found that the radiation tends to escape
through the gas rather than coherently lifting it, resulting in a
`steady-state' of turbulent gas boiling near the radiating bottom
surface.

\SD{} investigated the idea that the failure to lift the gas has to do
with the RT method. This is a valid concern, since the mean free paths
are, for the most part, resolved in the experiment, but FLD is
strictly only valid in the optically thick regime. They ran the
experiment with the \athena{} moment method RHD code, comparing the
FLD closure against the more accurate variable Eddington tensor (VET)
closure, which constructs the radiation flux vector on the fly in
every volume by sweeping the grid with short characteristics rays,
thus incorporating the contribution from all radiation sources and
absorbers. They found that the qualitative result is sensitive to the
closure used, with their FLD implementation giving a similar result as
found by \KT{}, while the VET version coherently lifts the gas out of
the frame. However, while the average horizontal velocity of the gas
is considerably higher with VET, the average optical depths and
radiation force on the gas are quite similar between the two methods:
the defining difference appears to be that the radiation force with
VET is just enough to lift the gas while with FLD it is just below
what is needed. The reason, the authors conclude, is that as the gas
is being lifted, the FLD closure tends to create chimneys in the gas
though which most of the radiation escapes, and hence the force is
enough to get the gas moving and forming those chimneys, but the
radiation never builds up sufficiently to evacuate the gas.

The M1 closure can be seen as an intermediate approach between those
of FLD and VET: instead of simply following the energy gradient as in
FLD, M1 stores locally the bulk direction of radiation, keeping some
`memory' of where it was emitted. However, the directionality of
radiation from multiple sources tends to mix locally, creating an
artificial diffusion which should be more or less absent with the VET
closure, provided good angular resolution in the VET ray-sweeping
scheme. We should therefore expect our results with M1 to lie
somewhere between those of FLD and VET, though a priori it is unclear
exactly where. Nonetheless, the quantitative results using the FLD and
VET closures in \SD{}, in terms of effective optical depths, radiation
force, and even gas velocities, lie within a fairly narrow margin,
making this a good test case for our implementation. We thus repeat
the test from \SD{} and validate our implementation by comparing our
results to theirs.

The setup of the experiment is as follows: the simulation box is a 2-D
square of height $L_{\rm box}=1024 \, h_{*}$, where
$h_{*}=2 \times 10^{15}$ cm is the scale height for the initial gas
density profile. The box is resolved by $2048^2$ cells, and the
resolution is fixed, i.e. we do not use adaptive refinement. The
physical resolution and box height is identical to that of \SD{},
while the box width, constrained by the square geometry of \ramses{},
is twice as large.  A layer of gas is placed at the bottom of the box,
and given an exponential density profile with distance from the
bottom, $\rho(h) = \rho_{*} \exp(-h/h_{*})$, where\footnote{Since the
  experiment is in 2-D, the units for density and column density
  should be ${\rm g \, cm}^{-2}$ and ${\rm g \, cm}^{-1}$,
  respectively. However, following \KT{} and \SD{}, we use 3-D units
  in the description for this experiment.}
$\rho_{*}=7.1 \times 10^{-16} \, \gcc$, resulting in a column density
of $\Sigma = 1.4 \ \gc$. Following \SD{}, we add fluctuations to the
initial gas density profile, of the form
\begin{align} \label{rho_fluctuations.eq}
  \frac{\partial \rho}{\rho} = 0.25 \left( 1 \pm \chi \right) 
  \sin\left( 2\pi x/ L_{\rm box} \right),
\end{align}
where $\chi$ is a random number in the range $\left[-0.25,
  0.25\right]$.  The initial gas profile is floored at a minimum
density of $10^{-10} \rho_{*}$, and the gas is given a homogeneous
initial temperature of $T_{*}=82$ K. The only non-adiabatic source of
heating and cooling for the gas is the dust-radiation interaction,
\begin{align} \label{r_d_coupling.eq}
  \frac{\partial e}{\partial t} =
  -\frac{\partial E}{\partial t} =
  \kP \rho \left(\cred E - c \ar T^4\right).
\end{align}

The bottom boundary of the box emits a radiation flux of
$F_{*}=1.03 \times 10^4 \, \eflux$
($2.54 \times 10^{13} \, L_{\odot} \, \rm{kpc}^{-2}$), and the box is
initialised to contain an upwards radiation flux of the same
magnitude, with $\cred E=F_{y}=F_{*}$ and $F_{x}=0$, and thus a
radiation temperature
\begin{align}
  T_{\rm r *}=\left( \frac{F_{*}}{c \ar} \right)^{1/4} =T_{*}.
\end{align}
The radiation is coupled to the gas via Rosseland and Planck opacities
which, vitally to the mechanics of this experiment, are functions of
the gas temperature:
\begin{align} \label{kappas.eq}
  \kP = 0.1 \left( \frac{\Tgas}{10 \, \rm{K}} \right)^2 \ \ccg, 
  \nonumber \\ 
  \kR = 0.0316 \left( \frac{\Tgas}{10 \, \rm{K}} \right)^2 \ \ccg.
\end{align}
These opacity functions originate from \KT{} and are approximately in
agreement with dust models at $T \la 150$ K \citep{Semenov:2003hk}.
Given the initial temperature, $T_{*}=82$ K, the initial Rosseland
opacity is $\kappa_{\rm R *} = 2.13 \, \ccg$.

The radiation force is countered by a homogeneous gravitational
acceleration field pointing downwards, of magnitude
$g=1.46 \times 10^{-6}\, \acc$.  The local competition between
downwards gravity and upwards radiation pressure is described by the
Eddington ratio,
\begin{align} \label{fEdd.eq}
  f_{\rm E} = \frac{\fyrad}{g \rho},
\end{align}
where $\fyrad$ is the vertical radiation force,
\begin{align} \label{fRad.eq}
  \fyrad = \frac{\kR \rho F_{y}}{c} + \frac{1}{3}\nabla \Etrap.
\end{align}
Given the initial conditions, the Eddington ratio is $f_{\rm
  E,*}=0.5$, so the radiation initially cannot lift the gas against
the opposing force of gravity. However, the gas is optically thick to
the radiation with an initial optical depth of, from bottom to top,
\begin{align}
  \tau_{*}=\kappa_{\rm R *} \Sigma = 3.
\end{align}
Thus, the radiation can be trapped and accumulated by the layer of
optically thick gas, which boosts the radiation temperature. Due to
the coupling in \Eeq{r_d_coupling.eq}, this in turn heats the gas,
which may via \Eeq{kappas.eq} increase $\kR{}$ to the extent that
$f_{\rm E}>1$. This of course requires efficient trapping of the
radiation, which is the vital factor that in the end decides whether
the gas is lifted or not.

It should be noted that {\it trapping} here not only refers to our
method for trapping radiation in regions where the optical depth is
unresolved, but also to radiation which may be free-streaming in
optically thin gas, but is trapped bouncing back and forth between the
confinements of optically thick shells. We do apply our method of
trapping photons \emph{inside} gas cells of unresolved mean free path,
which turns out to be relevant only to the early lift of gas, as we
shall see in the following analysis.

The box is periodic in the horizontal direction, both for the
radiation and matter. For the matter content, the bottom of the box is
reflective, allowing no escape or entry of gas, and Dirichlet boundary
conditions, i.e. fixed values, are applied to the top, with
$\rho=10^{-13} \, \rho_{*}$, $\Tgas=10^{-3}\, T_{*}$, and zero
velocity, in pressure balance with the initial conditions, and
allowing easy escape of upwards moving gas. For the radiation, we also
apply Dirichlet boundary conditions at the top, with zero flux and
energy density. The bottom boundary needs to emit radiation vertically
at the rate $F_{*}$. We accomplish this by solving the GLF intercell
flux function (\Eq{GLF.eq}) to give an intercell flux of
$F_{1/2}=F_{*}$ at the interface between each cell at the lower box
boundary and its ghost neighbour, with the additional requirement that
the ghost region cell has a photon flux of ${\bf F}_{0}=(0,F_{*})$.
This gives a radiation energy density for the ghost cell of
\begin{align} \label{ghost_cE.eq}
  \cred E_{0} = F_{*} - F_{y, 1} + \cred E_{1},
\end{align}
where the subscripts $0$ and $1$ refer to the ghost cell and the
boundary cell inside the computational domain, respectively. As with
all other tests presented in this paper, we use here the GLF intercell
flux function for calculating the photon advection between cells. We
tried as well with the HLL intercell flux function, which is better at
maintaining the directionality of radiation (see \RRT{}), though
photon trapping is strictly not supported with it (see comment in
\Sec{solver.sec}). Using HLL results in slightly more efficient early
lift of gas than with the GLF function, but eventual convergence
towards the same qualitative situation at the end of the run.

We follow the evolution of the system for $200 \, t_{*}$, where
$t_{*}=h_{*}/c_{*}$ is the characteristic sound crossing time, and
$c_{*}=\sqrt{\kb T_{*}/(\mu m_{\rm H})} = 0.54 \, \kms$ is the
characteristic sound speed.  We run the experiment using a reduced
light speed of $\cred=3 \times 10^{-3} \, c$, which is more than two
orders of magnitude faster than $c_{*}$ (and much faster than any gas
velocities attained in the experiment). We \emph{start} the experiment
at a full light speed and converge exponentially towards $\cred$ over
$3 \times 10^4$ RHD time-steps. We do this specifically to capture the
sudden and short lived pile-up of trapped photons by the gas which is
accumulated mostly in the bottom layer of cells. This only affects the
acceleration of gas in the initial few $t_{*}$, compared to running at
$\cred$ for the whole experiment.  We have run as well with a factor
of ten lower value for $\cred$, which gives a very similar evolution,
implying light speed convergence around the default value. In all the
results presented here, we use the relativistic corrections described
in the Appendix, but note that they have no visible effect on the
results.

To illustrate the effect and importance of photon trapping, we present
results from two \ramsesrt{} runs, one with and one without photon
trapping. The run without trapping uses $\cred$ for the whole run,
without the initial decrement from the full light speed, as this has
no effect without the trapping mechanism which is responsible for the
initial pile-up of radiation. Also, since the run without trapping has
much less initial vertical acceleration of gas, it has half the box
width (and height) as the one with trapping activated, while keeping
the same physical resolution, i.e. the box has a height of
$L_{\rm box}=512 \ h_{*}$ and is resolved by $1024^2$ cells.

\begin{figure}
  \centering
  \includegraphics
  {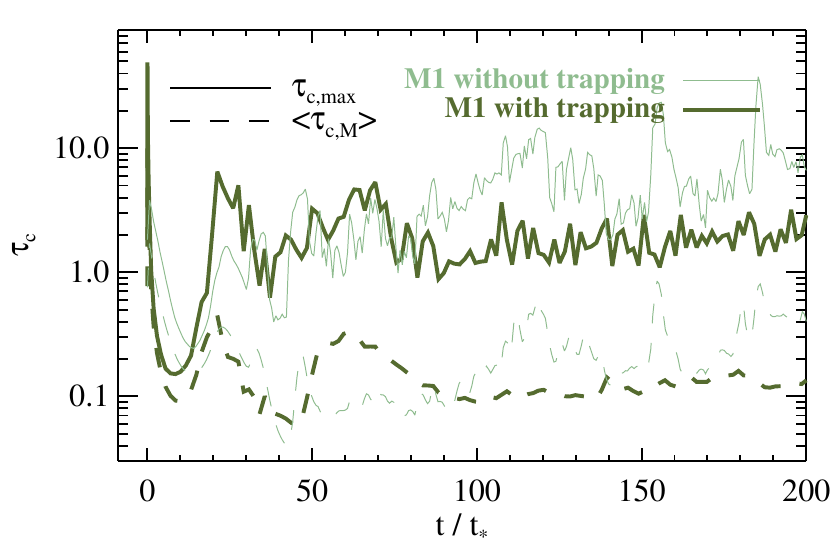}
  \caption{\label{davis_tau.fig}Maximum (solid) and mass-weighted
    average (dashed) cell optical depths in the gas levitation
    test. The thin bright-green curves show a run without radiation
    trapping, while the thick dark-green curves show the main run with
    radiation trapping. The high optical depths of cells indicate that
    the diffusion limit is somewhat relevant in this experiment,
    especially at the very start of the runs ($t\la 5 t_{*}$), where
    most of the gas mass is in the diffusion limit ($\tauc \ga 1$).}
\end{figure}

\Fig{davis_tau.fig} shows the evolution of cell optical depths,
$\tauc$. Focusing first on the run without trapping (light green
curves), we find that the mass weighted average and maximum cell
optical depths start at $\left<\tau_{c, M}\right>\approx2$ and
$\tau_{\rm c, max}\approx 4$, respectively, showing that the mean free
paths are unresolved at the start of the run, which implies that the
diffusion limit, and thus the photon trapping mechanism, is relevant
at the start. The cell optical depths quickly decline in value as the
gas rises from the bottom and becomes more diffuse, such that the mean
free path becomes better resolved. For the remainder of the run the
average cell optical depths are mostly well below unity, although
there always remain cells with large optical depths. With trapping
turned on (darker green curves), the optical depths start well above
the values from the non-trapping run, due to the larger concentration
of photons that now accumulates in the optically thick gas, which
leads to higher gas opacity via \Eeq{kappas.eq}. However, once the gas
starts to lift, the cell optical depths are reduced to smaller values
than in the non-trapping run, as a result of the diffusive pressure of
the trapped photons.  After the experiment has reached a turbulent
equilibrium state, around $100 \, t_{*}$, the opacities are
consistently lower than when trapping is not used.

\begin{figure*}
  \centering
  \includegraphics[width=1\textwidth]
  {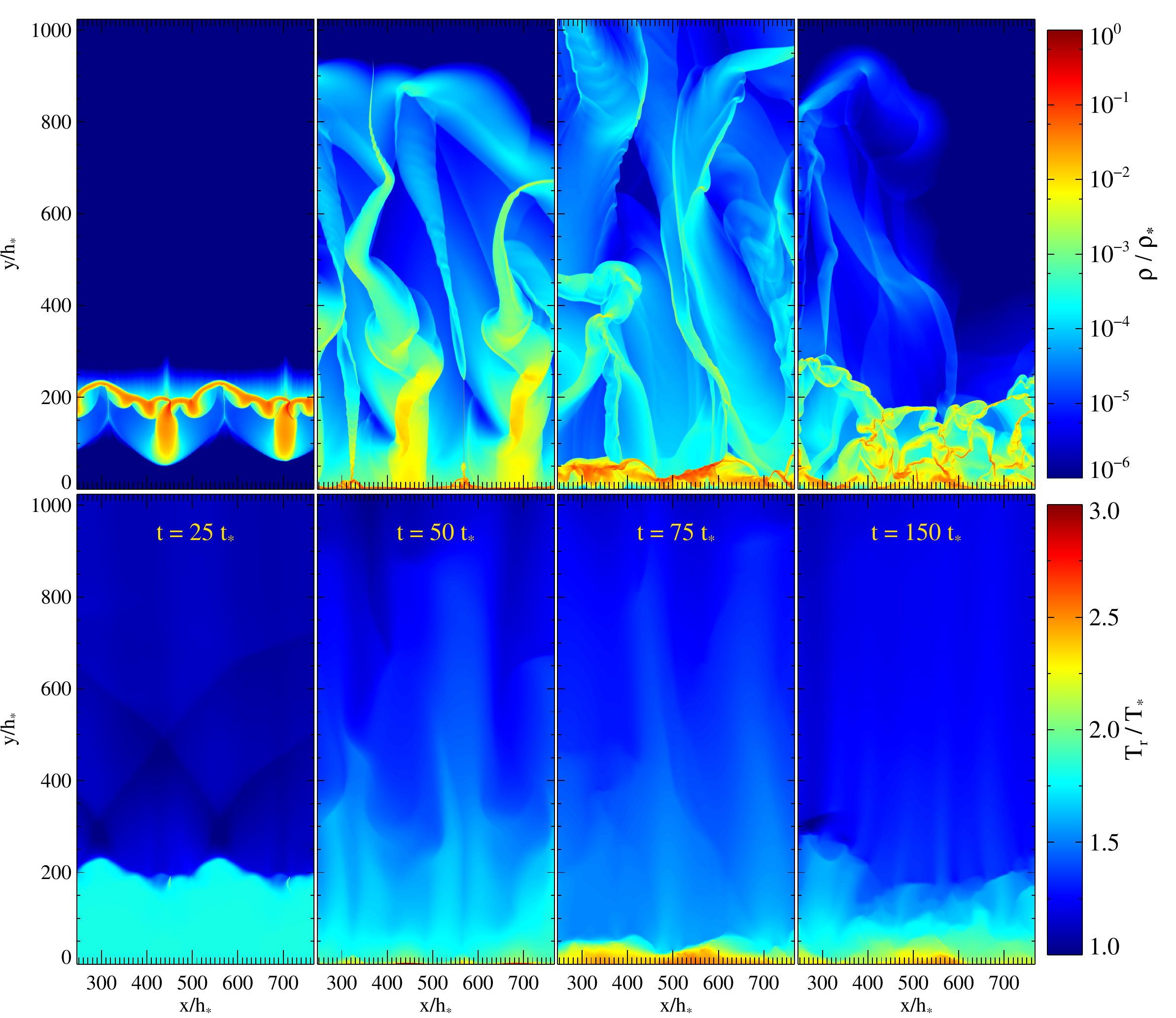}
  \caption{\label{davis_rho.fig}Maps of the gas density (upper row)
    and radiation temperature (lower row) in selected snapshots from
    the gas levitation experiment. We show the full height of the box,
    but to fit the maps on the page, we show only half of the width,
    along the center.}
\end{figure*}

\Fig{davis_rho.fig} shows maps of gas density and radiation
temperature at different snapshots of the run with photon
trapping. The evolution is qualitatively similar to the results in
\SD{}, and we see the same features of filamentary gas concentrations
interspersed with more diffuse `chimneys' through which the radiation
escapes to the top of the box. Visual inspection of the gas density
and radiation temperature suggests that the results fall in between
those of FLD and VET in \SD{} (their figures 3, 4, and 5). Focusing on
the gas densities, the gas is initially levitated quite efficiently,
even more so than in either FLD or VET, due to the strong initial
trapped photon pressure (a point which we will revisit later). About
$1\%$ of the total mass is ejected from the top of the box in the
first upwards burst of gas. The rest of the gas drops back to the
bottom, to $\la 200 \, h_{*}$, where it is kept turbulent by the
competition between radiation pressure and gravity.  Unlike with VET,
the gas is not coherently lifted beyond
$h\approx 500 \, h_{*}$\footnote{The VET simulation is restarted with
  an extended box height at $t=80 \, t_{*}$, when the gas approaches
  the upper boundary, and the gas is approaching the (new) upper limit
  at $h \approx 2048 \, h_{*}$ when the run is stopped at
  $\approx 150 \, t_{*}$.}. It settles to eventually occupy similar
heights as in the FLD results, where it is concentrated below
$\approx 200 \, h_{*}$ at $t=150 \, t_{*}$. The radiation temperature
maps show trapped radiation beneath coherent layers of gas, which
extends quite high initially, but is kept at much lower heights once
the gas breaks up due to Rayleigh-Taylor instabilities.

The first two density maps from the left ($t=25$ and $50 \ t_{*}$)
contain a conspicuous perfectly vertical feature at
$x\approx 575 \ h_{*}$. This gas is flowing downwards in a thin
stream, which is limited in thickness only by the cell width. The
horizontal forces on the gas stream are negligible for some time, and
thus, guided by the grid alignment, the stream can maintain this
perfect shape from $t\approx 22 \ t_{*}$ until it is destroyed by
laminar gas flows at $t=64 \ t_{*}$. No other such numerical features
appear in the simulation.

\begin{figure}
  \centering
  \includegraphics
  {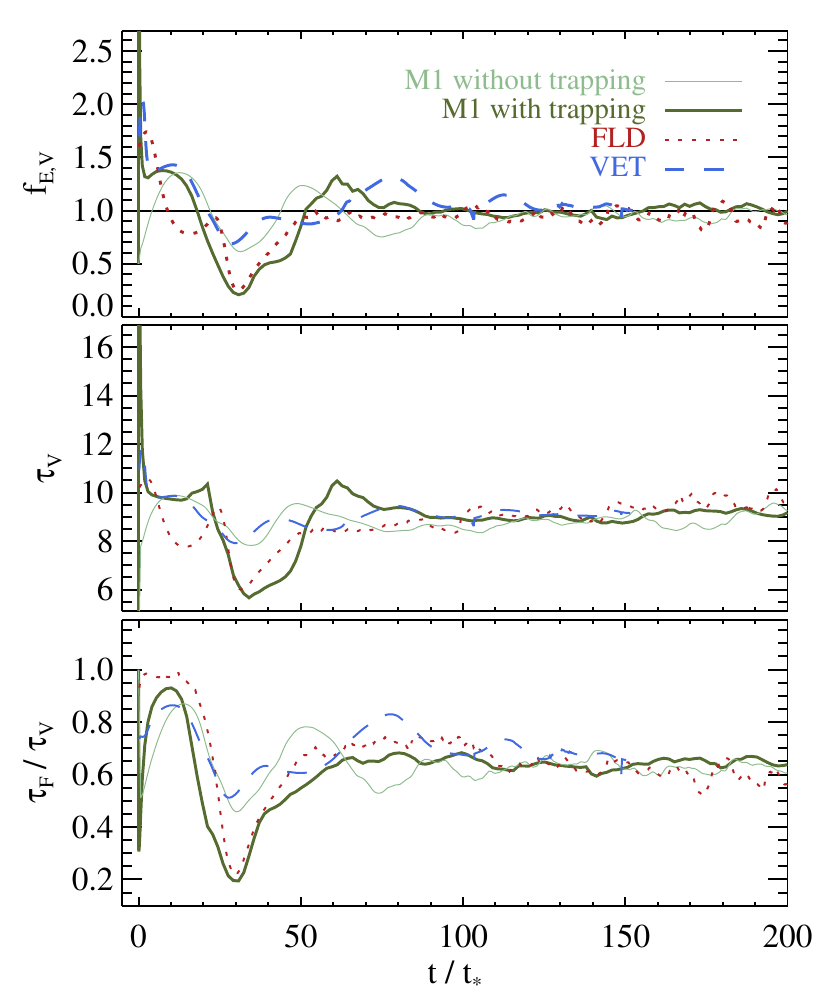}
  \caption{\label{davis_edd.fig}Comparison of gas levitation test for
    \ramsesrt{} with and without trapping (light green and darker
    green curves respectively), and for the \athena{} code, taken from
    \SD{}, using FLD (red) and VET (blue). {\bf Top panel}: Eddington
    ratio $f_{\rm E}$ ($=0.5$ at $t=0$) between the upwards force of
    radiation pressure and the downwards force of gravity. {\bf Middle
      panel}: average volume weighted optical depth along lines of
    sight from the bottom to the top of the box ($=3$ at $t=0$).  {\bf
      Bottom panel}: ratio between the flux weighted and volume
    weighted average optical depths ($=1$ at $t=0$). All plots show
    strong similarity between the different methods and
    codes. Comparison of the \ramsesrt{} results with and without
    trapping reveals that the diffusion limit is important at the
    beginning of the run, where a pile-up of radiation results in very
    strong optical depth and in turn a strong radiation force.}
\end{figure}

\begin{figure}
  \centering
  \includegraphics
  {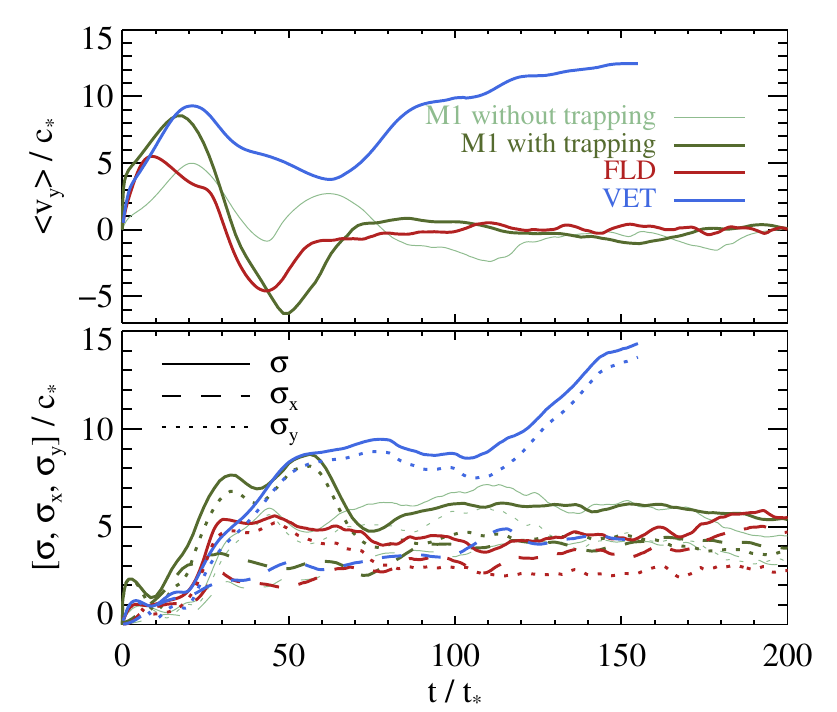}
  \caption{\label{davis_sigmas.fig}Gas velocity comparison in gas
    levitation test, for \ramsesrt{} with and without trapping (light
    green and darker green curves respectively), and for the \athena{}
    code from \SD{}, using FLD (red) and VET (blue). \textit{Top
      plot:} mass weighted mean vertical velocity. \textit{Bottom
      plot:} mass weighted velocity dispersions. The plots show good
    comparison between \ramsesrt{} and \athena{}, but the \ramsesrt{}
    results are more in line with the ones obtained with FLD than
    VET. The main effect of photon trapping in \ramsesrt{} can again
    be seen in the faster early acceleration due to the combination
    trapped photon pressure and the higher opacity of the gas that
    results from the trapped photons (\Eq{kappas.eq}).  }
\end{figure}

In figures \ref{davis_edd.fig} and \ref{davis_sigmas.fig} we compare
our results directly to those of FLD and VET from \SD{} (courtesy of
Shane Davis). The top plot in \Fig{davis_edd.fig} shows the volume
averaged Eddington ratio,
\begin{align} \label{fEV.eq}
  f_{\rm E,V} = \frac{\left< \fyrad \right> }{\left< g \rho \right> }.
\end{align}
This ratio expresses the competition between radiation pressure and
gravity, with $f_{\rm E,V}>1$ when radiation pressure has the upper
hand. By construction, $f_{\rm E,V}=f_{*}=0.5$ at the start of the
run. The middle plot shows the volume averaged optical depth from
bottom to top,
\begin{align} \label{tauV.eq}
  \tauV = \Lbox \left< \kR \rho \right>.
\end{align}
The evolution of this quantity is closely linked to $f_{\rm E,V}$
through that of $\kR$, which sets both the optical depth and the
strength of the radiation pressure. The bottom plot shows the ratio of
the photon flux-weighted mean optical depth,
\begin{align} \label{tauF.eq}
  \tauF = \Lbox \frac{\left< \kR \rho F_{y} \right> }{F_{y}},
\end{align}
to $\tauV$.

We first focus on the effect of photon trapping in the \ramsesrt{}
runs (\Fig{davis_edd.fig}, light and dark green curves). With photon
trapping turned on, there is an almost instantaneous rise from the
initial values, $f_{\rm E,V}=0.5$ and $\tauV=3$, quickly followed by a
steep decline in both. This early evolution is absent in the
non-trapping run, which just shows a gradual and much slower initial
rise for both quantities. The steep rise is due to the sudden buildup
of trapped photons in the bottom layer of cells, which increases
$\kR$. This results in a strong force from the diffusive radiation,
which quickly pushes the gas upwards. The rapid diffusion of the gas
in turn leads to a rapid decrease of $\kR$, and some of the trapped
radiation escapes upwards, reducing the opacity and the radiation
push. With trapping turned off, there is much less initial buildup of
radiation, and the initial push is gentler. In the long run, ignoring
the evolution in the first $\approx 10 \ t_{*}$, the evolution
with/without trapping, however, is quite similar.

The same can be said if we compare the \ramsesrt{} results to those
from \SD{}. The results agree quite well overall, showing similar
early reaction and then settling on similar semi-constant values of
$\fEdd$, $\tauV$, and $\tauF/\tauV$. In the early reaction phase,
$t \la 75 t_{*}$, the results in places resemble an interpolation
between the FLD and VET results, in line with our argument that M1 is
an intermediate approach between FLD and VET.  

The run with photon trapping very quickly reaches peaks of $\fEdd=10$
and $\tauV=32$ at $0.023 \, t_{*}$, which disappear rapidly as the gas
starts moving. We do not show these peaks in the plots in
\Fig{davis_edd.fig} for the sake of not stretching out the y-axes. The
magnitude of the peaks depends on the speed of light, which is the
reason why we start the trapping run with a full speed of light and
converge to $\cred$ in the first $\approx 3 \times 10^4$ time
steps. We verified in hydrodynamically static runs (i.e. with RT
turned on but the HD turned off) that an equilibrium is reached with
constant values of $\fEdd=10.7$ and $\tauV=32.5$, regardless of the
speed of light. The important differing factor is simply the time it
takes to reach that equilibrium, which with reduced light speed
becomes longer than the duration of the peak.

This rather large discrepancy in optical depth from the FLD and VET
implementations at early times demands further investigation to
justify our ballpark numerical value.  If we assume, for the sake of
simplicity, that all the gas is initially placed in a single
horizontal cell layer\footnote{This is a good approximation:
  $25-60 \%$ of the column density is initially in the bottom layer of
  cells, depending on the sinusoidal and random fluctuations.}, we can
derive an expression for the equilibrium value of the cell optical
depth, $\tauc$, at which the upwards flux from the cell equals
$F_{*}$. In the framework of M1 using the GLF intercell flux, with
photon trapping, such an equilibrium is met when
\begin{align}\label{eqiulibrium_M1_cond.eq}
\cred E_s = \left[1-\exp{(-\frac{2}{3\tauc})}\right]\cred E = 2 F_{*},
\end{align}
where $E_s$ is the $\tauc$-dependent streaming photon density
(\Eq{E_stream_mod.eq}). We can then combine the relation $\tauc=\kR
\Sigma$, \Eeq{kappas.eq} describing $\kR(T)$, and the relation
between radiation temperature and radiation energy, yielding
\begin{align}\label{eqiulibrium_M1.eq}
  \tauc(E) = 3.16 \times 10^{-4} \ \frac{\ccg}{\rm K^2} \ \Sigma 
  \ \sqrt{\frac{\cred E}{ca}},
\end{align}
assuming $\Trad=\Tgas$. Substituting \Eeq{eqiulibrium_M1.eq} into
\Eeq{eqiulibrium_M1_cond.eq}, and using $\Sigma_{*}$, then gives an
equilibrium condition that can be solved for $\tauc$, which yields a
median velue of $\tauc=27$, in fair agreement with our peak optical
depth of $32$. Allowing for the maximum fluctuation amplitude in
$\Sigma_{*}$ gives an upper limit of $\tauc=67$, and looking at
\Fig{davis_tau.fig}, we find that the maximum initial values for
$\tauc$ are within this limit.

With FLD we can make a similar estimate. Here the equilibrium
condition is
\begin{align}\label{eqiulibrium_FLD_cond.eq}
  \frac{\cred E}{3 \tauc} = F_{*},
\end{align}
and again using \Eeq{eqiulibrium_M1.eq} gives the same median and upper
limit for $\tauc$ as in the photon trapping framework. While these
simplified estimates do not predict the exact equilibrium value of the
optical depth, they demonstrate that the high initial peak reached in
our run is indeed plausible.

We now turn our attention to the gas velocities. The upper panel in
\Fig{davis_sigmas.fig} shows the ratio of the mass-weighted mean
(i.e. bulk) vertical velocity and the characteristic sound speed,
while the lower plot shows velocity dispersions in the gas
(i.e. turbulence). Without trapping, the M1 results show relatively
weak initial upwards acceleration of the gas, followed by a drop, a
bounce, and then an turbulent equilibrium state, with the velocity
dispersions well below the constantly rising ones of VET, but somewhat
above those of FLD. With trapping turned on, there is a much more
dramatic initial acceleration of gas, even stronger than that of VET,
which we already attributed to the strong initial buildup of trapped
radiation in and below the bottom layer of gas. This is followed by a
very strong deceleration and drop back to the bottom of the box, which
is even stronger than with FLD. The strong drop is likely due to the
reduced speed of light: the incoming radiation flux cannot keep up
with filling the growing `bubble' between the bottom of the box and
the rising layer of gas, and as a result the radiation pressure
deflates as the gas lifts. At the same time, radiative Rayleigh-Taylor
instabilities fragment the gas, allowing the radiation to escape, and
the gas falls hard back to the bottom. However, it also bounces back,
and eventually reaches a turbulent state quite similar to the
non-trapping run, and to FLD, though the velocity dispersions are
stronger than with FLD.

\begin{figure}
  \centering
  \includegraphics
  {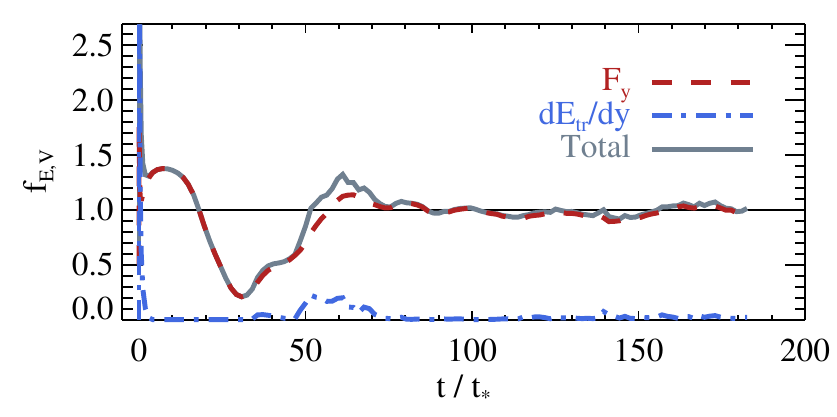}
  \caption{\label{davis_edd_contr.fig}Contributions, in the M1
    levitation test with trapping, to the total Eddington ratio
    (grey), from the free-flowing photon flux (red) and the diffusion
    pressure from trapped photons (blue). The diffusion pressure is
    important, but only at the very start of the run where almost all
    the gas mass is concentrated in one row of cells at the bottom of
    the box.}
\end{figure}

We finally illustrate, in \Fig{davis_edd_contr.fig}, the relative
contributions to the average Eddington ratio $f_{\rm E,V}$ (gray) from
the free-streaming photon flux, $\frac{\kR \rho F_{y}}{c}$ (red) and
from the trapped photon diffusion pressure $\frac{1}{3}\nabla \Etrap$
(blue). As suggested by the previous plots, the diffusion pressure
dominates strongly during the first few $t_{*}$, but is more or less
negligible for the remainder of the run.

Summarising this final test, we repeated with \ramsesrt{} the gas
levitation experiment described in \SD{}, in which FLD and VET
closures were used for solving the moment equations of RT. We ran the
same setup as described therein, modulo differences in the initial and
boundary conditions required by the different methods.  With FLD, the
bottom boundary condition requires that
\begin{align} \label{FLD_bottom.eq} \frac{c\lambda}{\kR
    \rho}\frac{\partial E}{\partial y} = F_{*},
\end{align}
where $\lambda$ is the flux-limiter that limits the speed of radiation
transport to the speed of light. With VET, the comoving radiation flux
in the bottom boundary ghost zone is set to $F_y=F_{*}$ which is quite
similar to the boundary condition we apply with M1, but they also add
a `diffusion limit' correction to the flux, enhancing it according to
the optical thickness of the layer of cells just above the
boundary. We need not apply any such correction, since the trapping of
photons automatically takes care of the diffusion limit. However, the
similar early evolution suggests that the correction made in VET is
valid, and that the diffusion limit is indeed mostly relevant in the
very bottom layer of gas cells.

All in all, our results using the M1 closure agree well with the other
closures, though they are qualitatively more similar to FLD than VET:
while both M1 and FLD manage to build up, after $50 \ t_{*}$, a
quasi-hydrostatic extended gas layer, VET still continues to evacuate
gas at a significant rate. In light of this, since the M1 closure does
not follow the gradient of radiation energy as the FLD closure does,
the difference between the fate of the gas with different closures is
likely to have a more nuanced explanation than just the FLD closure
tending to magnify radiative Rayleigh-Taylor instabilities. It is
non-trivial to read much in terms of physics into those differences,
especially since it remains to be seen how far the gas can levitate
with VET before reaching a turbulent equilibrium state, and whether
this state eventually resembles the results with FLD and M1.

While we cannot point out specifics in the other implementations which
could affect the experiment results, we can point out two factors
which might affect our own results. One is the reduced speed of
light. While our convergence tests that change the speed of light by a
factor of a few in each direction give very similar results, it is
possible that the results would be quite different if we used the real
speed of light, or a value close to it. Indeed we have seen that the
early acceleration of the gas is quite sensitive to the speed of
light, so is likely the relatively strong deceleration, and the same
may indeed apply later in the experiment. Possibly the gas can
spontaneously form a coherent layer that efficiently traps the
radiation. In such a scenario, the radiation builds up faster with an
increasing speed of light, and with a low speed of light the trapping
layer of gas may be destroyed by gravity ahead of the radiation
buildup, essentially keeping the gas from being lifted. Another factor
is the limitation of the M1 closure in dealing with multiple
sources. In the case of efficient trapping, the radiation essentially
bounces between the gas layer and the bottom of the box, and in such a
case the M1 closure may create an overtly diffusive radiation field
that tends to blow holes in the trapping layer of gas.

There are also limitations to the setup of this experiment, which
ultimately are probably more severe than the implementation details
mentioned, e.g. the lack of resolution in the initial setup, the close
competition between gravity and radiation, the monogroup approach, and
the lack of a third dimension.

In conclusion, and regardless of the physical limitations, this last
test gives support in favour of the robustness of the new additions to
\ramsesrt{}, as we test all the new aspects of the code,
i.e. radiation pressure, radiation-temperature coupling, radiation
trapping, and relativistic corrections (though the last factor turns
out to have no effect on the results). The results using \ramsesrt{}
are very similar to those obtained by FLD and VET in terms of the
evolution of the Eddington ratio between the forces of radiation and
gravity, the volume averaged optical depth, and the ratio between the
flux averaged and volume averaged optical depths. The early
acceleration of the gas is quite similar to the VET case, but instead
of continuing to lift, the gas drops back to the bottom and reaches a
turbulent equilibrium state, with velocity dispersions in-between
those of FLD and VET.


\section{Conclusions} \label{Conclusions.sec} We have presented
several important modifications to the RHD implementation in
\ramsesrt{}. Previously, as described in \RRT{}, the implementation
focused on the interaction of photons and gas via photo-ionisation and
the associated gas heating. In the current work, three features were
added:
\begin{itemize}
\item Multi-scattered IR radiation, which is coupled to the evolution
  of the gas/dust temperature. A vital ingredient here is the novel
  treatment of radiation diffusion in a medium where the mean free
  path is unresolved, by partitioning the radiation into sub-groups of
  {\it trapped} and {\it streaming} photons. In the optically thick
  limit, the method accurately reproduces the results of flux-limited
  diffusion (FLD), but has the great advantage over FLD that
  free-streaming photons are much more accurately modelled, and that
  photons can `adaptively' alternate between trapped and
  free-streaming, depending on the local properties of the gas.
\item Relativistic $v/c$ corrections to the implementation of
  dust-coupled radiation, accounting for Doppler effects and hence the
  work done by the radiation on the gas.
\item Momentum transfer from radiation to gas, allowing for realistic
  modelling of the effects of radiation pressure, both direct pressure
  from ionising radiation, and from reprocessed multi-scattered
  radiation.
\end{itemize}

We used a series of test to validate our new additions. These included
a morphological assessment of a radiation field produced by the M1
closure around a galaxy disk (\Sec{jiang.sec}), a test of
dust-absorbed radiation in a homogeneous optically semi-thick medium,
where we compared to a full RT solution (\Sec{gonzalez.sec}), tests of
direct ionising radiation pressure in an initially homogeneous gas
around a luminous young stellar population (\Sec{dir_press_test.sec}),
a qualitative resolution convergence test for photon trapping in a
resolved versus unresolved optically thick gas (\Sec{diff_test.sec}),
quantitative tests of radiation diffusion in optically thick gas, with
a radiation flash in 2-D (\Sec{2d_diff_test.sec}), and a constant
radiation source in 3-D (\Sec{3d_diff_test.sec}), and, finally, a 2-D
test of the competition of gravity and multi-scattering IR radiation
where we compared our results in terms of average optical depths,
Eddington ratios, bulk gas velocities, and turbulence, against
previously published results with the \athena{} code, from
\cite{Davis:2014jl}. With the tests, we can demonstrate a robust
treatment in \ramsesrt{} of the interaction of radiation and gas via
photoionisation heating, direct pressure from ionising radiation, dust
heating, and momentum deposition by multi-scattering photons.

There are limitations to the RHD approach that we use in
\ramsesrt{}. As discussed in both this work and \RRT{}, the M1 moment
method which we employ has problems in dealing with situations of
overlapping radiation from different sources, especially in between
those sources. We have presented demonstrations of this particular
limitation, but we argue that even if the radiation is not always
propagated to full quantitative precision, it is qualitatively robust,
and generally adequate in relevant astrophysical scenarios. Another
limitation of the code is that while it does offer a multi-frequency
approach, it is quite crude, with only a handful of frequency bins
realistically attainable in standard simulations.  However,
\citep{Mirocha:2012iw} have shown that as few as four bins of
(ionising) radiation, if optimally placed in the frequency range, can
eliminate frequency resolution errors to high precision, and other
factors, such as resolution, likely become more limiting in studying
the effects of radiation feedback on galaxy evolution.

We will follow up on this work with RHD simulations to study the
effects of radiation feedback from stars and AGN on galaxy evolution,
morphology, and outflows, on cosmological, galactic, and ISM scales.

The \ramsesrt{} implementation, including all the new features
described here, is publicly available, as a part of the \ramses{}
code.\footnote{\url{https://bitbucket.org/rteyssie/ramses}}

\section*{Acknowledgements}
We thank Oscar Agertz, Jeremy Blaizot, Benoit Commercon, Yohan Dubois,
and Yan-Fei Jiang for helpful discussions. Special thanks go to Shane
Davis and Matthias Gonzalez for kindly sharing data and information,
to Joop Schaye for many suggestions and corrections to the manuscript,
and the referee, John Wise, for constructive comments.  We acknowledge
the organisers and participants of the workshop \emph{Gravity's Loyal
  Opposition}, held at KITP in Santa Barbara April 14th - July 3rd
2014, at which many components of this work came together. The work
was funded by the European Research Council under the European Union’s
Seventh Framework Programme (FP7/2007-2013) / ERC Grant agreement
278594-GasAroundGalaxies, and the Marie Curie Training Network
CosmoComp (PITN-GA-2009-238356). The simulations were mostly performed
using the DiRAC Data Centric system at Durham University, operated by
the Institute for Computational Cosmology on behalf of the STFC DiRAC
HPC Facility (www.dirac.ac.uk). This equipment was funded by BIS
National E-infrastructure capital grant ST/K00042X/1, STFC capital
grant ST/H008519/1, and STFC DiRAC Operations grant ST/K003267/1 and
Durham University. DiRAC is part of the National E-Infrastructure. We
also used the HPC resources of CINES under the allocation
2011-c2011046642 made by GENCI (Grand Equipement National de Calcul
Intensif), and computing resources at the CC-IN2P3 Computing Center
(Lyon/Villeurbanne - France), a partnership between CNRS/IN2P3 and
CEA/DSM/Irfu.

\bibliography{ref}

\newpage

\appendix

\section{Relativistic Corrections to the RHD
  Equations} \label{vc1.app}

We describe briefly the RHD equations, taking into account $v/c$ terms
that were missing in this paper so far, which represent relativistic
Doppler effects between the rest frames of the gas and the
radiation. These equations are derived form the classical textbook on
RHD, \cite{Mihalas:1984vm}.

We now distinguish between the radiation energy expressed in the gas
comoving frame, noted $E_0$, from the radiation energy in the lab
frame, noted $E$ in the main text. We also define the radiation flux
vector in the comoving frame as ${\bf F}_0$, and the lab frame
radiation flux ${\bf F}$.  The gas total energy is defined as usual by
\begin{align}
E_{\rm gas}=\frac{1}{2}\rho v^2 + e,
\end{align}
where we recall $\rho$ and $v$ are the gas density and speed,
respectively, and $e$ is the gas internal thermal energy.

We now add $v/c$ terms to the radiation momentum equations
(\ref{mom_E.eq}-\ref{mom_F.eq}), neglecting only $(v/c)^2$ terms
\citep[see][page 423]{Mihalas:1984vm}.
\begin{align}
\frac{\partial E}{\partial t} + \nabla \cdot {\bf F} 
   &= \kappa \rho \left( c a T^4 - \cred E 
   + {\bf v} \cdot \frac{1}{c}{\bf F} \right), \label{mixE.eq}\\
\frac{\partial {\bf F}}{\partial t} + \cred^2 \nabla \cdot 
   {\mathbb D} E 
   &= \kappa \rho \cred \left(
   - {\bf F} + {\bf v} a T^4 
   + {\bf v} \cdot \frac{\cred}{c}{\mathbb D} E \right). 
   \label{mixF.eq}
\end{align}
Note that $\lambda = (\kappa \rho)^{-1}$ is the frequency averaged
mean free path {\it computed in the comoving frame}. Doppler effects
are therefore only accounted for up to $v/c$ in the previous explicit
form, and the radiation variables are still in the lab frame. This
formulation is therefore referred to as {\it the mixed frame
  equations}.

We find it convenient to re-express these equations using the comoving
radiation variables, when coupled to the thermochemistry.  For this,
we use the Lorentz transform up to first order in $v/c$ to compute
comoving variables as a function of the lab frame variables.  We have
\citep[][page 417]{Mihalas:1984vm}:
\begin{align}
E_0 &= E - \frac{2}{\cred c}{\bf v} \cdot {\bf F}, \label{lorentz_e} \\
{\bf F}_0 &= {\bf F} - {\bf v} \cdot \frac{\cred}{c} E 
   \left( {\mathbb I} + {\mathbb D} \right) \label{lorentz_f}.
\end{align}
Injecting these relations into the mixed frame equations
(\ref{mixE.eq}-\ref{mixF.eq}) leads to the  form
\begin{align}
\frac{\partial E}{\partial t} + \nabla \cdot {\bf F} 
   &= \kappa \rho \left( c a T^4 - \cred E_0 \right)
   -  {\bf v} \cdot \frac{\kappa \rho}{c} {\bf F}, \label{Ecom.eq}\\
\frac{\partial {\bf F}}{\partial t} 
   + \cred^2 \nabla \cdot {\mathbb D} E
   &= \kappa \rho \cred {\bf F}_0 
   + {\bf v} \frac{\kappa \rho \cred}{c} 
   \left( c a T^4 - \cred E \right). \label{Fcom.eq}
\end{align}
The source terms are now easier to interpret: the first term on the
right-hand side of the energy equation is the classical radiation and
matter coupling term in the comoving frame. The second term is equal
to minus the work of the radiation force in the lab frame. In the
radiation flux equation, the first term is the radiation force in the
comoving frame, while the second one is a purely relativistic term
usually identified as a {\it frame dragging effect} between matter and
radiation. The gas energy and momentum equations
(\ref{lte_egy.eq}-\ref{fluid_mom.eq}, ignoring gravity and other
heating/cooling processes) are modified accordingly and are written
using a globally strictly conservative form
\begin{align}
\frac{\partial E_{\rm gas}}{\partial t} 
   + \nabla \cdot \left( {\bf v} (E_{\rm gas}+P) \right) 
   &= \kP \rho \left( \cred E_0 - c a T^4 \right)
   -  {\bf v} \cdot \frac{\kP \rho }{c} {\bf F}, \\
\frac{\partial \rho {\bf v}}{\partial t} 
   + \nabla \cdot \left( \rho {\bf v} \otimes {\bf v} 
   + P{\mathbb I} \right) 
   &= \frac{\kR \rho}{c} {\bf F}_0 
   - {\bf v} \frac{\kR \rho}{c^2} \left( c a T^4 - \cred E \right). 
\end{align}

We directly exploit this form of the RHD equations in our numerical
implementation, by adding each contribution in a classical operator
splitting approach.

\section{Trapped versus streaming photons in a mixed frame
  framework} \label{vc2.app} In order to deal with extremely opaque
conditions, for which the mean free path, $\lR=(\kR \rho)^{-1}$, is
much smaller than the grid spacing $\Delta x$, we have developed in
Section~\ref{trapping2.sec} a trapped/streaming radiation approach
that properly captures the diffusion limit, even if one does not
resolve the mean free path. This method was presented without taking
into account the relativistic corrections discussed in the previous
section. We now consider both the comoving and the lab frame, and our
trapped photons are assumed to be isotropic {\it in the comoving
  frame}. This means that ${\bf F}^0_t=0$ and, to first order in
$v/c$, one has from Eqs. (\ref{lorentz_e}-\ref{lorentz_f}):
\begin{align}
E_t=E^0_t{\rm ,}~~~{\mathbb P}_t=\frac{E^0_t}{3}{\mathbb I}
~~~{\rm and}~~~
{\bf F_t}=\frac{4}{3}\frac{\cred}{c}E^0_t{\bf v},
\end{align}
where we now express the comoving variables with a `$0$' superscript
rather than a subscript. We split the radiation energy into trapped
and streaming components $E=E^0_t+E_s$, using the decomposition of
Section~\ref{trapping2.sec} based on the local cell optical depth. The
total radiation energy equation (\Eq{Etstot.eq}, ignoring the
$\dot{E}$ source term) then becomes
\begin{align}
  \nonumber
  \frac{\partial E^0_t}{\partial t} + \frac{\partial E_s}{\partial t} 
  + \nabla \cdot \left( {\bf F_s} 
  + \frac{4}{3}\frac{\cred}{c} E^0_t{\bf v}
  \right) =\\
  \kP \rho \left( c a T^4 - \cred E^0_t - \cred E^0_s \right)  
  - {\bf v} \cdot \frac{\kP \rho}{c}{\bf F}, \label{Eradtot_mix.eq}
\end{align}
and the total radiation flux equation (\Eq{t-s-flux.eq}) becomes
\begin{align}
  \nonumber
  \frac{\partial {\bf F_s}}{\partial t} + \frac{\cred^2}{3} \nabla E^0_t 
  + \cred^2 \nabla \cdot \left( {\mathbb D} E_s \right)=\\ 
  - \kR \rho \cred {\bf F}^0_s 
  + {\bf v} \frac{\kR \rho \cred}{c} \left( c a T^4 - \cred E \right).
  \label{Fradtot_mix.eq}
\end{align}
In the diffusion regime, we would like to recover \Eeq{Fstream.eq} in
the comoving frame, i.e.
\begin{align}
{\bf F}^0_s \simeq - \frac{\cred \lR}{3} \nabla E^0_t.
\end{align}
In order to enforce our scheme to satisfy this limit when the cell
size is large compared to the mean free path, we exploit our GLF flux
function (\Eq{GLF_formal.eq}) and we fix the streaming to trapped
photon ratio by
\begin{align}
E^0_t = \frac{3\tauc}{2}E_s.
\end{align}
We then solve for the streaming photon energy and flux variables in
Eqs. (\ref{Eradtot_mix.eq}-\ref{Fradtot_mix.eq}) using
our mixed frame M1 Godunov solver
\begin{align}
\frac{\partial E_s}{\partial t} + \nabla \cdot {\bf F}_s
   &= - \kP \rho \cred E^0_s  - {\bf v} \cdot
   \frac{\kP \rho}{c}{\bf F}_s, \\
\frac{\partial {\bf F}_s}{\partial t} 
   + \cred^2 \nabla \cdot {\mathbb D} E_s
   &= - \kR \rho \cred {\bf F}^0_s +
{\bf v} \frac{\kR \rho \cred}{c} \left( c a T^4 - \cred E \right).
\end{align}
The total radiative force is decomposed into a streaming and a
trapped component as before,
\begin{align}
\frac{\kR \rho}{c}{\bf F}=\frac{\kR \rho}{c}{\bf F}_s
-\frac{1}{3}\frac{\cred}{c}\nabla E^0_t.
\end{align}
The work of the radiation force (with a minus sign) is decomposed
between the streaming and the trapped photon energy equation. For the
latter, we solve the trapped part of the radiation energy equation
(\ref{Eradtot_mix.eq}), namely
\begin{align}
\frac{\partial E^0_t}{\partial t} 
+ \nabla \cdot \left( \frac{4}{3}\frac{\cred}{c}E^0_t{\bf v} \right) 
= \kP \rho \left( c a T^4 - \cred E^0_t \right)  
+ {\bf v} \cdot \frac{1}{3}\frac{\cred}{c}\nabla E^0_t,
\end{align}
which can be re-written as the classical comoving radiation energy
equation
\begin{align}
\frac{\partial E^0_t}{\partial t} + \nabla 
\cdot \left(\frac{\cred}{c} E^0_t{\bf v} \right) 
+ \Prad \nabla \cdot {\bf v}
= \kP \rho \left( c a T^4 - \cred E^0_t \right),
\end{align}
where the trapped radiation pressure is
$\Prad=\frac{1}{3}\frac{\cred}{c}E^0_t$.  The gas momentum equation
(\ref{mom_trapped.eq}, ignoring the gravity term) is also modified
into
\begin{align}
\nonumber
\frac{\partial \rho {\bf v}}{\partial t} 
+ \nabla \cdot \left[ \rho {\bf v} \otimes {\bf v} 
+ (P+\Prad){\mathbb I} \right] = \\
 \frac{\kR \rho}{c}{\bf F}^0_s 
- {\bf v}\frac{\kR \rho}{c^2} \left( c a T^4 - \cred E \right),
\end{align}
as well as the gas total energy equation (\ref{lte_egy.eq}, ignoring
gravity and $\Lambda$),
\begin{align}
\nonumber
\frac{\partial}{\partial t}\left( E_{\rm gas}+E^0_t \right) 
+ \nabla \cdot \left[ {\bf v} (E_{\rm gas}+E^0_t+P+P^0_t) \right] =\\ 
\kP \rho \cred E^0_s + {\bf v} \cdot \frac{\kP \rho}{c}{\bf F}_s.
\end{align}
We now see quite clearly that in very optically thick regions, where
$E_s \ll E^0_t$, the streaming photons energy and flux can both be
ignored and the previous set of equations just becomes a classical HD
system with two pressure and energy components (gas and trapped
radiation), that can be solved with a multi-fluid Godunov scheme.  We
incorporate the trapped energy radiation energy and pressure into all
components of the fluid solver, as in \cite{Commercon:2011eq}.

\section{A full RT solver} \label{fullRT.app}
In \Sec{gonzalez.sec} we compare \ramsesrt{} results to a full
radiative transfer calculation, which we will now describe.

The full RT solver takes a `bulldozer' approach in solving the full
radiative transfer equation, (\ref{RT.eq}), in the four-dimensional
space $(x,y,\phi,\theta)$, where the first two dimensions are location
and the latter two are the standard solid angle, with $\phi$ the angle
from the $x$-axis in the $xy$-plane and $\theta$ the angle from the
normal vector to the $xy$-plane. The four-dimensional space is
discretised into a four-dimensional grid $(i,j,k,\ell)$, with a total
number of elements $N_x \times N_y \times N_{\phi} \times N_{\theta}$,
where the $N$s denote the number of bins in each dimension. Each grid
element contains the radiation specific intensity $I(i,j,k,\ell)$ (in
a single group approach). The radiation energy density (energy per
unit volume) in a cell $(i,j)$ is retrieved by summing the specific
intensity over all angles:
\begin{align}
  E(i,j) = \frac{1}{c} \sum_{k=1}^{N_{\phi}} \sum_{\ell=1}^{N_{\theta}}
  I\left( i,j,k,\ell \right) \, \sin{\theta} \ \Delta \theta \, \Delta \phi,
\end{align}
where
\begin{align}
  \Delta \phi &= \frac{2 \, \pi}{N_{\phi}} \\
  \Delta \theta &=  \frac{\pi}{N_{\theta}}, \\
  \phi(k) &= (k-1.) \Delta \phi, \\
  \theta(\ell) &= (\ell-0.5) \Delta \theta.
\end{align}

The specific intensity is integrated on the whole grid, according to
\Eeq{RT.eq}, in discretised time-steps of length
$\dt=0.5\frac{\dx}{c}$. In each timestep, the specific intensity is
updated from $I^{t}$ to $I^{t+\dt}$ in three operator-split steps:
injection, advection, and scattering, which are performed as follows.

\subsection{Injection} \label{fullRT_inj.sec} This step corresponds to
solving \Eeq{RT.eq} with only the first term on the RHS, i.e.
\begin{align}
  \frac{1}{c}\frac{\partial I}{\partial t} = \eta.
\end{align}
Here, photons are simply added to $I(i,j,k,\ell)$ where appropriate.

In our \Sec{gonzalez.sec} test, no such injection inside the box
boundaries is in fact needed. Here, it suffices to initialise the
boundary conditions such that the correct flux is emitted from the
left side. For all but the left boundary, the ghost cells, i.e. static
cells just outside the box boundary, are initialised to zero radiation
intensity, while for the left-side ghost cells we set
\begin{align}
  I(0,j,1,\ell) = \frac{1}{2} 
  \ \frac{F_*}{\sin{\theta}\Delta \phi \Delta \theta},
\end{align}
for $j=(1,...,N_y)$ and $\ell=(N_{\theta}/2-1, N_{\theta}/2)$,
assuming even $N_{\theta}$.

\subsection{Advection} \label{fullRT_adv.sec}
Here, we solve \Eeq{RT.eq} over $\dt$ with only the advection term,
i.e.
\begin{align}
\frac{1}{c}\frac{\partial I}{\partial t} + 
{\bf n}\cdot \nabla I = 0.
\end{align}
First, fluxes are calculated across each intercell boundary inside the
grid (and at the grid boundaries). The $x$-fluxes are
\begin{align}
  f_x(i+\tfrac{1}{2},j,k,\ell) = c n_x I_{\downarrow}
  (i+\tfrac{1}{2},j,k,\ell),
\end{align}
where $n_x=\cos{\phi} \, \sin{\theta}$, and $I_{\downarrow}$ is the
downstream radiation intensity, i.e. 
\begin{align}
  I_{\downarrow}(i+\tfrac{1}{2},j,k,\ell) &= 
  \begin{cases}
    I(i,j,k,\ell) & \rm{\ if \ }  n_x > 0, \nonumber \\
    I(i+1,j,k,\ell), & \rm{\ otherwise.}   \nonumber
  \end{cases}
\end{align}
Likewise, the $y$-intercell fluxes are
\begin{align}
  f_y(i,j+\tfrac{1}{2},k,\ell) = c n_y I_{\downarrow}
  (i,j+\tfrac{1}{2},k,\ell),
\end{align}
where $n_y=\sin{\phi} \, \sin{\theta}$, and
\begin{align}
  I_{\downarrow}(i,j+\tfrac{1}{2},k,\ell) &= 
  \begin{cases}
    I(i,j,k,\ell) & \rm{\ if \ }  n_y > 0, \nonumber \\
    I(i,j+1,k,\ell), & \rm{\ otherwise.}   \nonumber
  \end{cases}
\end{align}

The radiation is then explicitly advected between cells, using the
intercell fluxes:
\begin{align}
  I'(i,j,k,\ell) &= I(i,j,k,\ell)
  + \frac{\dt}{\dx} \,  \\
     & \left[ \, f_x(i-\tfrac{1}{2},j,k,\ell) - f_x(i+\tfrac{1}{2},j,k,\ell)
       \right. \nonumber \\
     & + f_y(i,j-\tfrac{1}{2},k,\ell) - f_y(i,j+\tfrac{1}{2},k,\ell)
       \left. \right], \nonumber 
\end{align}
for each $i \in (1,...,N_x)$, $j \in (1,...,N_y)$,
$k \in (1,...,N_{\phi})$, $\ell \in (1,...,N_{\theta})$.

\subsection{Scattering} \label{fullRT_sc.sec}
In the final operator-split step in the full RT calculation, the
radiation is scattered isotropically. First, the radiation intensity
in each cell and over all angles is semi-implicitly `absorbed':
\begin{align}
  I''(i,j,k,\ell) = 
  \frac{I'(i,j,k,\ell)}{1+\dt \, \rho \kappa c}.
\end{align}  
Then these photons are emitted isotropically (i.e. scattered):
\begin{align} \label{final_update.eq}
  I^{t+\dt}(i,j,k,\ell) =  I''(i,j,k,\ell) +
  \frac{f_{\rm sc}(i,j)}{4 \pi},
\end{align}  
where $f_{\rm sc}$ is the scattered flux over the timestep,
\begin{align}
  f_{\rm sc}(i,j) = \sum_{k=1}^{N_{\phi}} \sum_{\ell=1}^{N_{\theta}} 
  \left[ I''(i,j,k,\ell) -  I'(i,j,k,\ell) \right]
  \sin{\theta} \Delta \phi \Delta \theta.
\end{align}

With \Eeq{final_update.eq}, the radiation specific intensities are
fully updated to time $t+\dt$, and now the sequence of operator
splitting steps (\ref{fullRT_inj.sec}-\ref{fullRT_sc.sec}) can be
repeated for consecutive time-steps.

\end{document}